\documentclass[onecolumn]{aastex631}

\PassOptionsToPackage{usenames}{color}

\usepackage{graphicx} 
\usepackage{comment}
\usepackage[normalem]{ulem}
\usepackage{hyperref}
\usepackage[usenames]{color}
\def\figspath{figures_pdf}
\DeclareGraphicsExtensions{.png,.pdf}
 
\definecolor{lightred}{rgb}{0.6,0.0,0.0}
\definecolor{lightgreen}{rgb}{0.1,0.5,0.1}
\definecolor{offwhite}{rgb}{0.7,0.7,0.5}

\definecolor{amethyst}{rgb}{0.6, 0.4, 0.8}
\definecolor{aqua}{rgb}{0.0, 1.0, 1.0}
\definecolor{offwhite}{rgb}{0.7,0.7,0.5}

\definecolor{rgreen}{rgb}{0.0, 0.26, 0.13}

\received{}
\revised{}
\accepted{}
\submitjournal{ApJ}

\begin{document}

\title{Evaluating Non-LTE Spectral Inversions with ALMA and IBIS}

\shorttitle{Spectral Inversions with ALMA and IBIS}
\shortauthors{Hofmann, Reardon, Milic, \textit{et al.}}

\correspondingauthor{Kevin P. Reardon}
\email{kreardon@nso.edu}

\author[0000-0002-5556-6840]{Ryan Hofmann}
\altaffiliation{DKIST Ambassador}
\affiliation{National Solar Observatory, Boulder, CO, 80303}
\affiliation{Department of Astrophysics and Planetary Sciences,
University of Colorado, Boulder 
CO, 80303}

\author[0000-0001-8016-0001]{Kevin P. Reardon}
\affiliation{National Solar Observatory, Boulder, CO, 80303}
\affiliation{Department of Astrophysics and Planetary Sciences,
University of Colorado, Boulder 
CO, 80303}

\author[0000-0002-0189-5550]{Ivan Milic}
\affiliation{National Solar Observatory, Boulder, CO, 80303}
\affiliation{Department of Astrophysics and Planetary Sciences,
University of Colorado, Boulder 
CO, 80303}

\author[0000-0003-0583-0516]{Momchil Molnar}
\altaffiliation{DKIST Ambassador}
\affiliation{National Solar Observatory, Boulder, CO, 80303}
\affiliation{Department of Astrophysics and Planetary Sciences,
University of Colorado, Boulder 
CO, 80303}

\author[0000-0001-5615-3581]{Yi Chai}
\affiliation{Center for Solar-Terrestrial Research, New Jersey Institute of Technology, Newark, NJ, 07102}

\author[0000-0002-2554-1351]{Han Uitenbroek}
\affiliation{National Solar Observatory, Boulder, CO, 80303}

\begin{abstract}
We present observations of a solar plage in the millimeter-continuum with the ALMA and in the Ca 8542 and Na 5896 spectral lines with the Interferometric BIdimensional Spectrometer (IBIS). Our goal is to compare the measurement of local gas temperatures provided by ALMA with the temperature diagnostics provided by non-LTE inversions using STIC. 
In performing these inversions, we find that using column mass as the reference height scale, rather than optical depth, provides more reliable atmospheric profiles above the temperature minimum and that the treatment of non-LTE hydrogen ionization brings the inferred chromospheric temperatures into better agreement with the ALMA measurements.
The Band 3 brightness temperatures are higher but well correlated with the inversion-derived temperatures at the height of formation of the Ca 8542 line core. The Band 6 temperatures instead do not show good correlations with the temperatures at any specific layer in the inverted atmospheres.
We then performed inversions that included the millimeter continuum intensities as an additional constraint. Incorporating Band 3 generally resulted in atmospheres showing a strong temperature rise in the upper atmosphere, while including Band 6 led to significant regions of anomalously low temperatures at chromospheric heights.
This is consistent with the idea that the Band 6 emission can come from a range of heights. The poor constraints on the chromospheric electron density with existing inversion codes introduces difficulties in determining the height(s) of formation of the millimeter continuum as well as uncertainties in the temperatures derived from the spectral lines. 

\end{abstract}

\keywords{Solar chromosphere (1479), Millimeter astronomy (1061), Radiative transfer (1335), Solar atmosphere (1477)}

\section{Introduction}
\label{Ch:intro}

In recent years, solar physics has gained a new tool for the study of the chromosphere---high-resolution observations in the millimeter wavelengths from the Atacama Large Millimeter/submillimeter Array \citep[ALMA,][]{2009IEEEP..97.1463W}.
The solar radiation in the wavelength regime of 1 to 10 mm is believed to mostly arise from the continuum emission in the chromosphere; see \citet[][]{2016SSRv..200....1W} for discussion.
Previous observations in this regime were performed with smaller arrays, and thus had limited resolution; one such example is the Berkeley-Illinois-Maryland Array (BIMA) \citep{Loukitcheva_2009}, which has a resolution of 10 arcseconds, sufficient to reveal the shape of magnetic networks, but incapable of resolving the finer chromospheric structures that are apparent at shorter wavelengths \citep{2016SSRv..200....1W}.
With ALMA, spatial resolutions of 1 arcsecond or better are now achievable, fine enough to resolve distinct features in the solar atmosphere on similar spatial scales as seen in imaging spectrographs operating at optical and ultraviolet (or other) wavelengths.

This makes it possible to now directly measure the electron gas temperature in the mid- to upper-chromosphere at smaller spatial scales.
\citet{2007A&A...471..977W} demonstrated that the contribution functions for several of ALMA's bands peak in distinct regions of the chromosphere, with increasing wavelengths sampling increasing heights.
In a similar analysis, \citet{2015A&A...575A..15L} synthesized ALMA observations from a 3D radiation-MHD simulation of a quiet-sun network, finding that millimeter observations at a resolution of 1 arcsecond or better can provide a reasonable measurement of the electron temperature in the solar atmosphere.
Thus, millimeter observations can provide a direct measurement of the temperatures at various, if ill-determined, heights in the solar chromosphere, though care should be taken to account for all relevant physics and observational biases, especially the effects of line-of-sight integration and limited spatial resolution on observed millimeter intensities, to ensure correct inferences \citep[][]{2020ApJ...891L...8M, 2021A&A...656A..68E}.

Temperatures in the chromosphere have traditionally been obtained indirectly through the inversion of spectral lines \citep{2016LRSP...13....4D}.
``Inversion'' here refers to any mathematical problem that uses the value of an integral to determine the expression being integrated.
In this case, the expression is the physical conditions within the solar atmosphere, and the effects thereof on radiation propagating outward, e.g. absorption, emission, scattering, polarization; the integral itself is a summation of these effects along the line-of-sight; and the value of that integral is the emergent intensity at a given wavelength, which is what is observed.
These inversion techniques have already been applied with great success in the photosphere, with spectral lines such as the \ion{Fe}{1} 6302 \AA\ line and inversion codes assuming local thermodynamic equilibrium (LTE) such as SIR \citep{1992ApJ...398..375R}, and recent years have seen the development and application of new codes such as STiC \citep{2019A&A...623A..74D} and DeSIRe \citep[][]{2022arXiv220202226R}, which are capable of treating multiple lines in non-local thermodynamic equilibrium (non-LTE), particularly the \ion{Mg}{2} h and k lines, the \ion{Ca}{2} H and K lines, and the \ion{Ca}{2} infrared triplet.

While spectral inversions have proved quite useful in inferring the properties of the solar atmosphere, they can be somewhat inconsistent in certain regimes: especially in non-LTE, inversions are an ill-constrained problem with several possible degeneracies and pitfalls.
One way to remedy this problem is to include mm observations as an additional observational constraint in the inversions.
This was first demonstrated by \citet{2018A&A...620A.124D}, who used a slice from a 3D r-MHD simulation to compute synthetic observations, including both chromospheric spectral lines and millimeter continuum intensities, invert them with STiC in various combinations, and compare the results, finding that inclusion of millimeter diagnostics should greatly improve the reliability of inversions in recovering temperatures in the low- to mid-chromosphere.
More recently, \citet{2020A&A...634A..56D} applied this technique to real observations of a plage in the Mg II h and k lines and the Band 6 continuum, finding that the inclusion of the millimeter continuum reveals the presence of pockets of hot and cool gas in the low chromosphere that are invisible to the UV lines.

In this work, we explore some of these questions by performing the first non-LTE inversions of spectral lines in the visible and near-infrared whose heights of formation significantly overlap with the formation region of the millimeter continua sampled by ALMA. We examine the preferred height scale for inversions including the solar chromosphere and the importance of including non-LTE hydrogen ionization into account in that process.
We compare the temperatures derived from this inversion process with the brightness temperatures derived from the simultaneous high-resolution observations at 1.2 and 3.0 mm. We also perform tests where we include the ALMA continuum intensities directly in the inversion process to see if this additional information helps derive representative temperature profiles in the atmosphere. We examine the effects of the possible extended formation ranges of the millimeter continuum, especially for the Band 6 measurements, and the effect of uncertainties in the electron density in the inversion outputs.


\section{Observations}
\label{Ch:Observations}

\begin{figure*}
    \includegraphics[width=\textwidth]{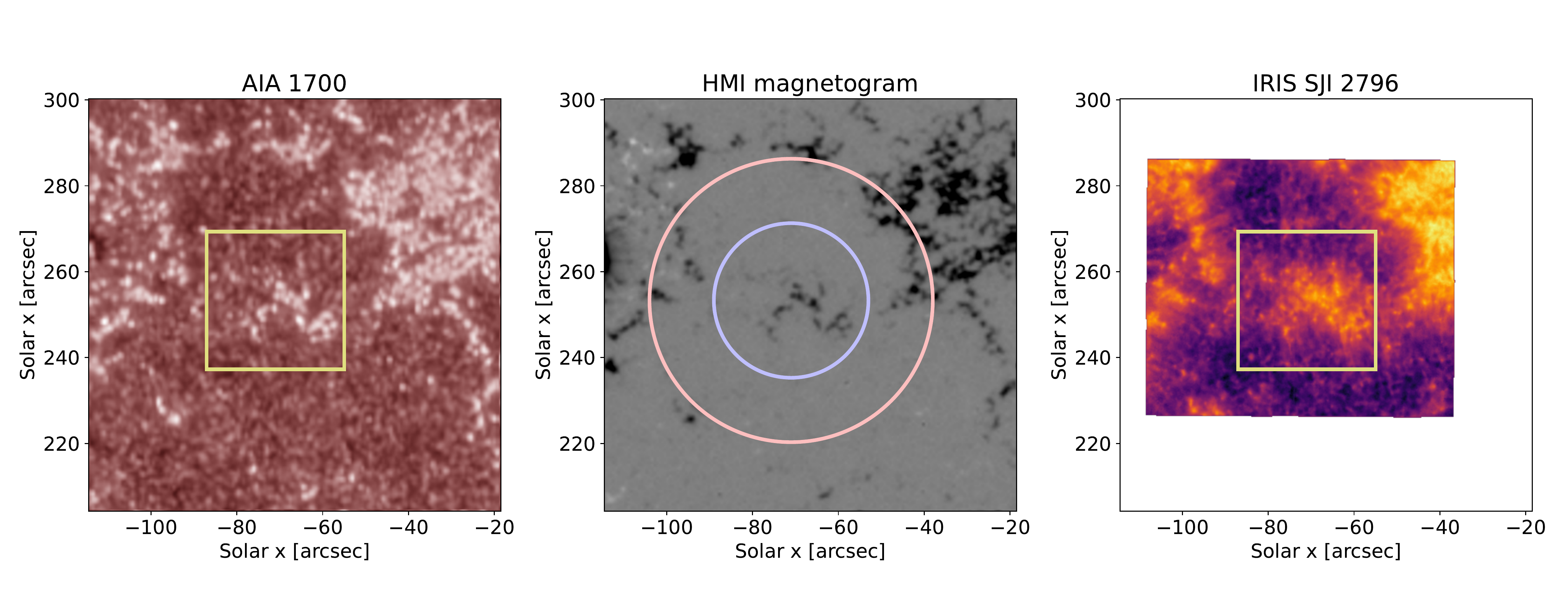}
    \caption{Context images of the observed field of view from the following instruments at 15:57 UT: \textbf{Left:} AIA 1700 {\AA} image; \textbf{Middle:} HMI LOS magnetogram, scaled (linearly) between -500 to 500 G.; \textbf{Right:} IRIS SJI image at 2796 {\AA}, averaged over one minute. The field of view of ALMA Band 6 is shown as the blue circle and the ALMA Band 3 field of view is shown as the red circle. The rectangle indicates the region of interest for which inversions were performed.}
    \label{fig:context}
\end{figure*}

We use observations obtained as part of coordinated program between ALMA and the Dunn Solar Telescope \citep[DST,][]{1991AdSpR..11..139D} on 2017 April 23.
At the DST, the Interferometric Bidimensional Spectrometer  \citep[IBIS, an imaging spectrograph][]{2006SoPh..236..415C,2008A&A...481..897R} observed the target region in several visible and near-IR spectral lines.
In addition, the IRIS \citep{2014SoPh..289.2733D} and Hinode \citep{2007SoPh..243....3K} satellites were co-pointing during the time of these observations, though these datasets were not included in our analysis.
Context images and magnetic field maps were available from SDO/AIA \citep{2012SoPh..275...17L} and SDO/HMI \citep{2012SoPh..275....3P}.

The observed target, shown in Figure \ref{fig:context}, was the leading portion of NOAA active region 12651, which was stable and showed no significant flaring activity. This same region, with observations take one day prior, was the one analyzed by \citep{2020A&A...634A..56D}. The center of the field of view (located at E04, N11, or $\mu = 0.96$, where $\mu$ is the cosine of the angle between the surface normal and the line of sight) could be classified as magnetic plage, with some quieter areas present in the southern portion of the IBIS and ALMA Band 3 fields of view (FOV). Even though the target was close to an active region, the average unsigned flux in the ALMA field of view was only about 20 G, as measured from the HMI magnetograms.
The leading spot of the active region was located about 50$\arcsec$ east of the target center, outside the area covered by IBIS and ALMA, but the penumbra of the spot did slightly extend into the edges of the IBIS field of view. 

\subsection{DST/IBIS Observations}
\label{sec:IBIS_obs}

\begin{figure}
    \includegraphics[width=\textwidth]{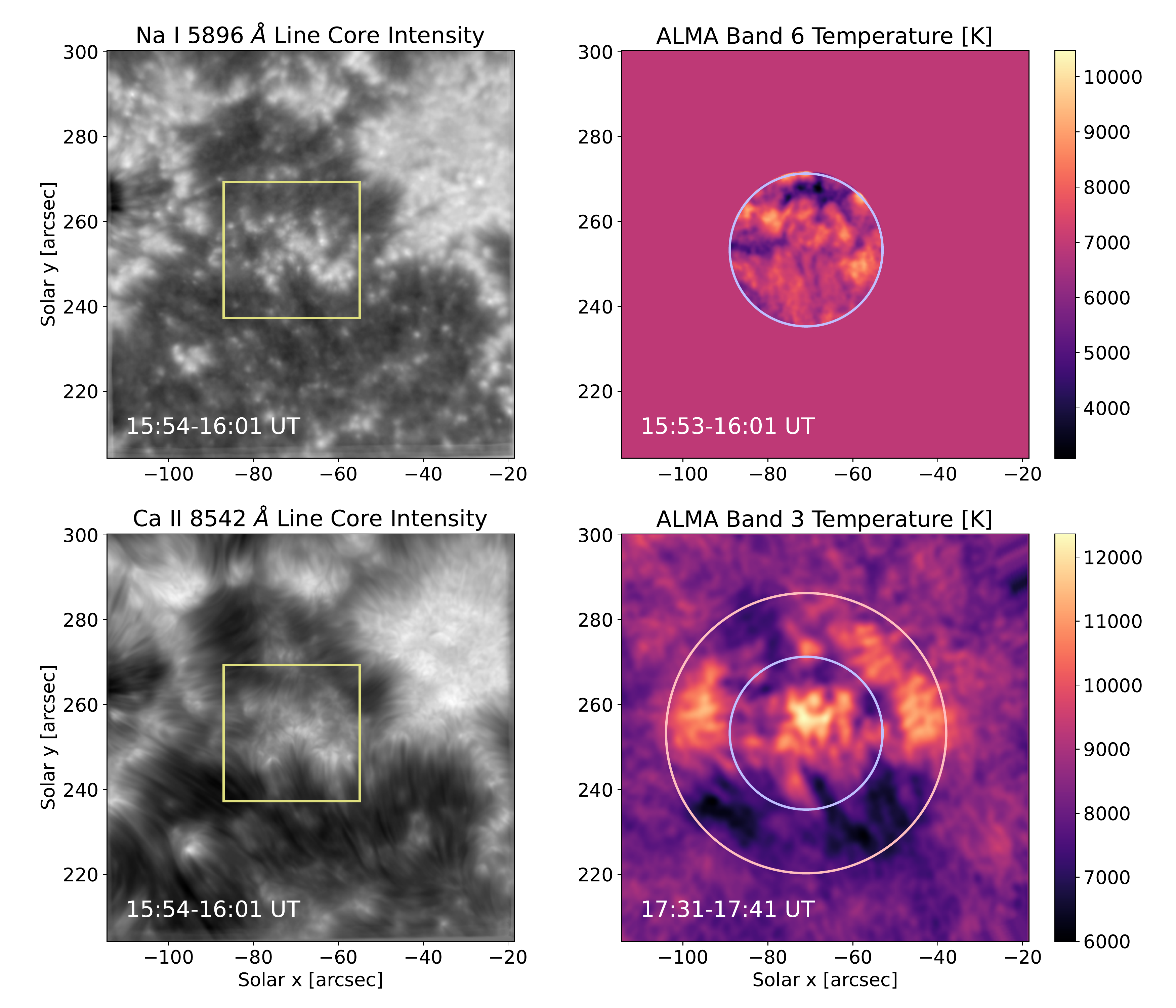}
    \caption{Context images from IBIS and ALMA, averaged over the indicated intervals. \textbf{Left panels:} Line minimum intensities observed with IBIS; the rectangle shows the region over which inversions were performed. \textbf{Right panels:} Brightness temperatures observed with ALMA (note later time for Band 3 observations); the circles indicate the approximate ALMA fields of view.}
    \label{fig:ibis_parameter_maps}
\end{figure}

We observed the target with IBIS from 15:13 to 19:06 UT in conditions of good to excellent seeing. 
The observations were taken in repeated scans of several different spectral lines (including \ion{Ca}{2} 8542 \AA\, \ion{Na}{1} D1 5896 \AA, and \ion{H}{1} 6563 \AA), in slightly different combinations or cadences. 
In this work, we use data taken during two series of scans acquired between 15:54--16:35 UT (hereinafter referred to as series 1554) and 17:25--18:12 UT (series 1725). 
The second series has been studied previously in \cite{2019ApJ...881...99M, 2021ApJ...920..125M}.
The cadence for the scans through all the observed lines was 10 and 16 seconds in the two series, respectively.
We perform our spectral inversions on a selected scan obtained during each series, in the time intervals 15:57:10--15:57:18 and 17:36:32--17:36:46.
We display the time-average of line-minimum-intensity images taken from 15:54-16:01 UT in the left column of Figure \ref{fig:ibis_parameter_maps}.

We applied linearity, dark correction, flat-fielding, and fringe-removal corrections to the IBIS data.
In order to correct for optical and atmospheric image distortions, we used the nearest-in-time HMI continuum intensity images as a reference in order to precisely map first the IBIS whitelight-channel images, and then the co-temporal spectral data onto a regular, fixed spatial grid (with an image scale of $0.096\arcsec$/pixel), effectively removing the bulk of the seeing distortions.
\footnote{using Rob Rutten's very capable software package available at \href{http://www.staff.science.uu.nl/~rutte101/Recipes_IDL.html}{his website}.}

\subsection{ALMA Observations and Processing}
\label{sec:ALMA_processing}

The ALMA data were obtained with the array in configuration C40-3, with a maximum baseline of 396 m (rather than the nominal 460 m).
The data analyzed here were snapshots obtained in Band 6 (1.2--1.3 mm, 229--249 GHz, 0.8\arcsec beam size) and Band 3 (2.8--3.3 mm, 92--108 GHz, 1.9\arcsec beam size) at 15:57 and 17:36 UT, respectively.
Even after the application of the standard interferometric calibration procedures using CASA \citep{2012ASPC..461..849P}, the ALMA images are still heavily influenced by the phase disturbances due to water vapor variations in Earth's atmosphere, which causes phase errors across the array and small-scale distortions on the images \citep{2017SoPh..292...87S}.
To counteract this effect, we apply the self-calibration technique, where an average (clean) image over some temporal interval is used to derive and remove the random phase variations in the individual images within that observing interval \citep[as described in more detail in][]{Chai_sunspot_2022}.
This allows us to minimize the atmospheric image distortions in a self-consistent manner that should not introduce significant artifacts.

The effective field of view of ALMA Band 3 (Band 6) images is only $\sim$ 60{\arcsec} (30{\arcsec}) in diameter (Figure \ref{fig:ibis_parameter_maps}, right), which is smaller than the large-scale solar structures.
To provide information on the more extended or background emission from the Sun, several single-dish observations that scan the full disk were taken nearly simultaneously with the interferometer array \citep{2017SoPh..292...88W}, albeit with a significantly lower spatial resolution.
Through the feathering process provided in CASA, we combined the two datasets covering the full range of spatial scales.
The full-disk image was normalized such that the brightness temperature in the central region of the disk (diameter 200 arcsec) in Bands 3 and 6 was set to 7300 K and 5900 K, as determined by \citet{2017SoPh..292...88W}, providing an absolute temperature accuracy of about 2-5\%.

\section{Inversions}
\label{Ch:Inversions}

The inversions were performed using the STockholm inversion Code \citep[STiC, see][]{2019A&A...623A..74D}, a regularized non-LTE inversion code based on the radiative transfer code RH, a multi-line, non-LTE spectral synthesis code \citep{2001ApJ...557..389U}.
The inversion code takes in the calibrated data, $I_{obs}(\lambda)$, and seeks a combination of plasma parameters, $\mathbf{p}$, that reproduce the observed spectra; in this case, the fitted parameters are temperature, line-of-sight (LOS) velocity, and microturbulent velocity, at a number of discrete depths in the model atmosphere.
For each iteration, the synthetic spectra, $I_{syn}$, are compared to $I_{obs}$ using the merit function:

$$\chi^2 (\mathbf{p}) = \frac{1}{n} \sum_{i=1}^N \left(\frac{I_{obs}(\lambda_i) - I_{syn}(\lambda_i,\mathbf{p})}{\sigma_i}\right)^2 + \sum_{j=1}^{N_p} \alpha_j r_j (\mathbf{p}).$$

Here, the first term is the chi-squared coming from the likelihood of the data given the model, and the second term is the sum of regularization terms for each model parameter, $n$ is the number of degrees of freedom and is equal to ($N-N_p)$, and $\sigma_i$ is the measurement uncertainty for intensity at $\lambda_i$.
Note that $\sigma_i$ can also be tuned to give certain wavelengths more weight than others; in our case, the line centers are given more weight in order to ensure a good fit at chromospheric heights.
If the fit is poor, i.e. $\chi^2$ is large, then the atmospheric parameters are adjusted according to the response of the emergent intensity, and the process repeats until convergence, upon which the code returns the best-fit atmospheric parameters and the corresponding synthetic spectra.

We inverted IBIS spectral scans of the \ion{Ca}{2} and \ion{Na}{1} lines together, in a similar manner to the multi-line inversions of \ion{Ca}{2} and \ion{Na}{1} performed by \cite{2022arXiv220203955D}, which required additional calibrations to the data.
The spatially dependent instrumental blueshift was removed by interpolating and resampling onto a uniform wavelength grid, which is necessary in order for the code to properly take into account the broadening profile of the instrument.
The wavelengths were weighted such that only the observed points contribute to the merit function.
To allow accurate inference of LOS velocities, the mean spectra were aligned with the FTS ATLAS \citep[][]{1993asps.book.....W, 1998assp.book.....W} to provide an absolute wavelength calibration.
Finally, the intensities were normalized to the disk center continuum.

An important aspect of the inversion process, especially in the non-LTE case, is the choice of the atomic models that incorporate radiative transitions other than the ones that are directly observed.
The iterative treatment of the populations of all these atomic levels makes non-LTE inversions numerically demanding and requires a balance between accuracy and simplicity.
We found that a 4-level sodium atom and a 6-level calcium atom provide a good compromise between computational speed and realism. 

\subsection{Height Scale Comparison}
\label{sec:height_scale}

\begin{figure}
    \centering
    \includegraphics[width=0.8\textwidth]{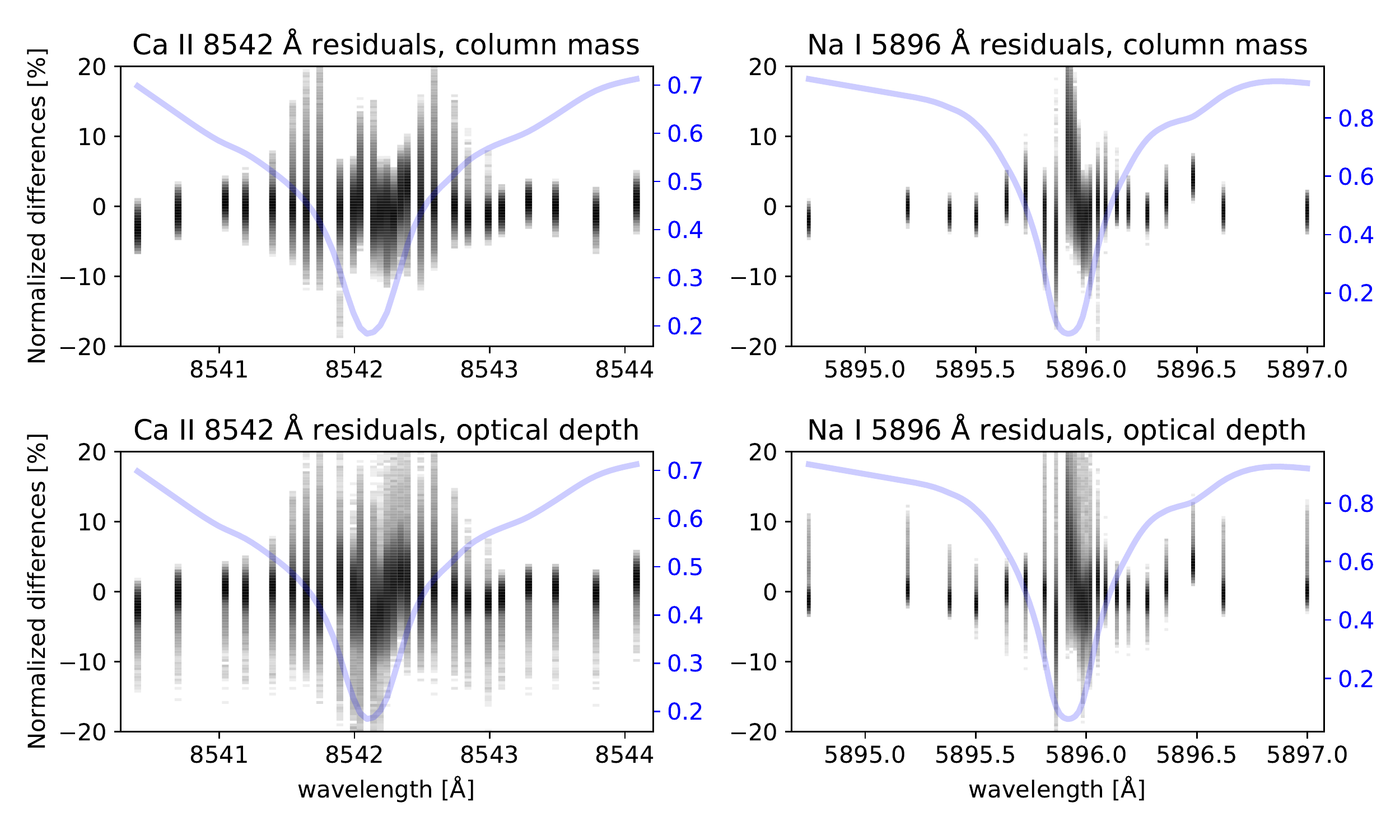}
    \caption{Histograms of differences between observed and inversion-fit profiles, normalized to the observed intensity at the local wavelength for series 1725 data, with inversions performed in column mass \textit{(top)} and optical depth \textit{(bottom)}, plotted for each of the sampled wavelength positions. 
    The average, continuum-normalized, observed spectra for the two spectral lines are overplotted in blue with intensities shown on the vertical axis on the right.}
    \label{fig:xvt_res}
\end{figure}

\begin{figure}
    \centering
    \includegraphics[width=0.8\textwidth]{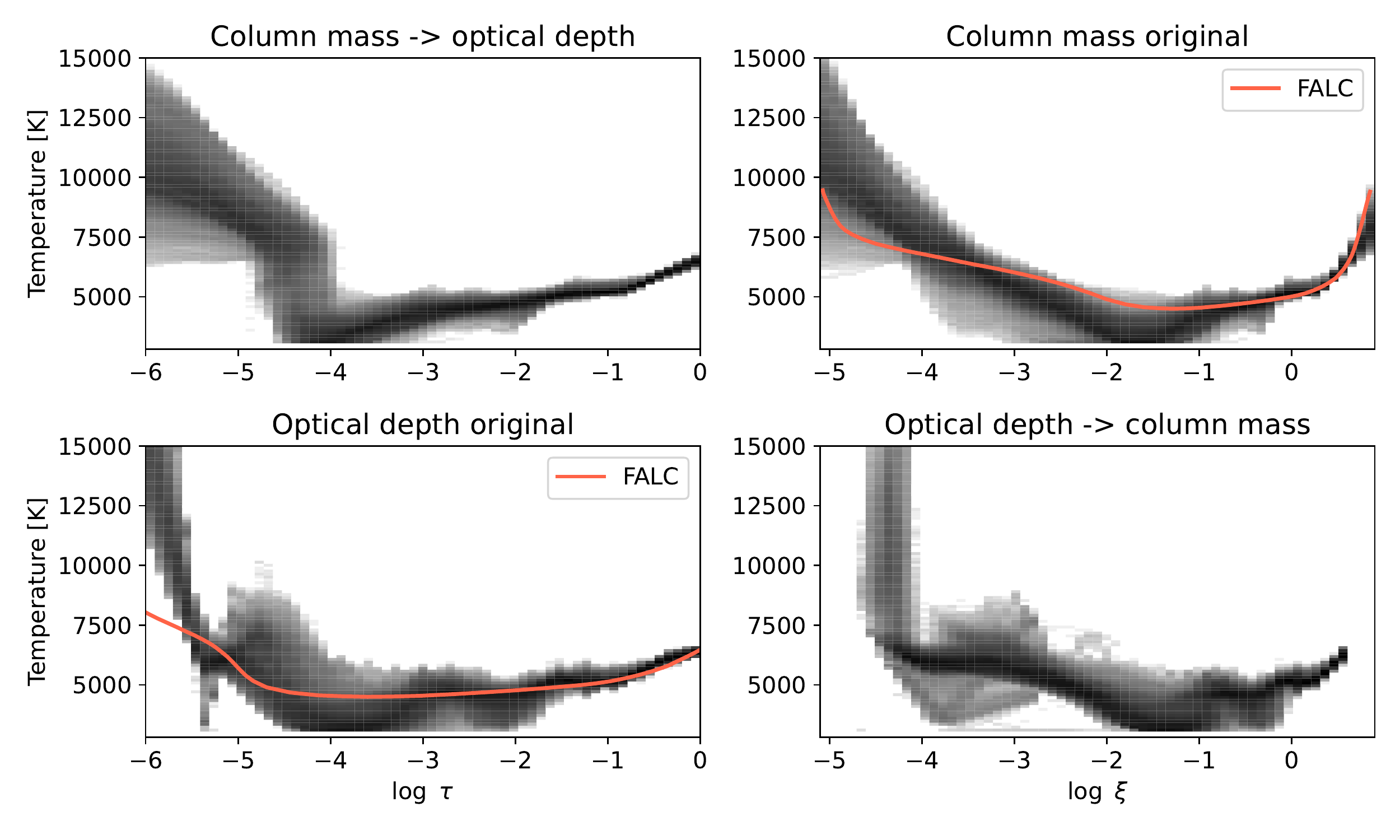}
    \caption{2-D histograms of inferred temperature profiles for series 1725 data, with inversions performed in column mass \textit{(top)} and optical depth \textit{(bottom)}. The top-right and lower-left panels show the profiles output from the inversions for the two different height scales, while the other two panels show the temperature profiles converted to the alternative height scale using the assumption of hydrostatic equilibrium. The sensitivity of the \ion{Ca}{2} 8542 \AA\ line doesn't well constrain the inversion above $\log\ \xi = -4.3$ or $\log\ \tau = -5.5$. Histogram intensities are log-scaled for clarity.}
    \label{fig:xvt_hist}
\end{figure}

Inversion codes such as STiC seek a height-dependent parameterization of the solar atmosphere, which results in a model with a large number of free parameters; namely, if the atmospheric model parameterizes $p$ physical properties over a range of $N$ height steps, then the model has $N \times p$ free parameters.
As a result, spectroscopic inversions are an ill-posed numerical problem with a high degree of degeneracy.
To make the problem more tractable, inversion codes parameterize each atmospheric parameter in terms of a smaller number of \textit{nodes}, specific heights in the atmosphere for which the numerical solutions are computed, and between which the atmospheric parameters are smoothly interpolated.
In order to accurately infer the temperature stratifications, it would nominally be ideal to use a relatively large number of nodes to recover as much vertical detail as possible, assuming sufficient computational resources.
Using many nodes, however, tends to lead to oscillatory solutions and poor convergence, which STiC circumvents through regularization, which penalizes oscillatory or divergent solutions.

The spectroscopic inversion processes described above does not recover the atmosphere along a fixed geometrical height scale.
Historically, inversions have always been performed using optical depth as the height scale, which has proved to work well for the photosphere where it gives a rather uniform vertical resolution.
However, optical depth scale greatly compresses the chromospheric structures: while the solar atmosphere up to the transition region spans roughly six orders of magnitude in both $\tau$ and $\xi$, the chromospheric temperature rise spans two orders in $\xi$ and less than one order in $\tau$ \citep[Figure 3]{2019A&A...623A..74D}; this can be a significant obstacle for high-fidelity inversions of chromospheric lines, especially when inverting for multiple chromospheric diagnostics simultaneously.
The STiC code introduced the possibility of using column mass ($\xi$) as the height scale.
This both simplifies the solution of the hydrostatic equilibrium and helps us to better resolve the chromospheric structures.
For inversions seeking to probe the structures and gradients of the chromosphere, column mass provides a more natural height scale, as the solar atmosphere is approximately hydrostatically stratified, and therefore, column mass is much more closely related to geometric height.

The nodes can be placed arbitrarily for either of these two height scales. To capture the chromospheric temperature gradients in the optical depth scale, however, would require using a large number of nodes in the regions above the (local) temperature minimum. Since the location and slope of the chromospheric temperature rise can vary significantly between profiles, it is difficult to have a optimal node placement in optical depth that consistently and accurately fits lines such as the calcium infrared triplet, which are sensitive to the location of this gradient. Because the chromosphere is not as compressed in column mass scale as in optical depth scale, we found that an equidistant or uniform node placement in column mass provided a much more robust inference of the atmospheric profiles.

We performed tests comparing inversions in optical depth and column mass, fitting the observed calcium and sodium lines together with simultaneous Band 3 continuum over the full plage region, in order to recover the temperature ($T$), line-of-sight velocity ($v_{\rm los}$), and microturbulent velocity ($v_{\rm turb}$).
For both the $\log\ \xi$ and $\log\ \tau$ inversions, we determined $[T,v_{\rm los},v_{\rm turb}]$ using [21,11,4] equidistant nodes covering the range of heights ($-5.1 < \log\xi < 0.9$ for column mass,  $-6.0 < \log\xi < 0.0$ for optical depth) from the lower photosphere to the upper chromosphere. The regularization weights chosen such that the penalty term in the merit function is of a similar magnitude as the $\chi^2$ term.
The quality of the fits (Figure \ref{fig:xvt_res}) for the inversions in $\log \tau$ was significantly worse overall than for $\log \xi$: while the final $\chi^2$ was overall similar for both (median $\chi^2$ for log $\tau$ inversion was 1.35 and for log $\xi$ inversion 1.44), the $\log\ \tau$ inversions had $\sim 10-100 \times$ more anomalous pixels that did not converge to $\chi^2 \leq 10$, and the inferred chromospheric temperatures tended to be less smoothly stratified in height, especially in the central plage regions. This can be seen by comparing the histogram of the temperature profiles recovered from the inversions for the two height scales, as shown in Figure \ref{fig:xvt_hist}. To further highlight the differences, we apply a conversion between these profiles into their corresponding alternative height scale based on the assumption of hydrostatic equilibrium, as shown in the upper left and lower right panels. For the column mass inversions, we used the per-pixel optical depth scales as computed by STiC, while for the optical depth inversions, we obtained the column mass from the total gas pressure: $\xi = p_{gas} / g$, where $g$ is the acceleration due to gravity.
This shows how the steep temperature rise into the chromosphere, which takes place smoothly between $-4 \le \log \xi \le -2$, is compressed into a much smaller interval of $-4.6 \le \log \tau \le -4$. This makes the inversion in optical depth particularly sensitive to the placement of the nodes such that they properly sample this steep change (in this height scale) in atmospheric parameters.

\begin{figure}
    \centering
    \includegraphics[width=\textwidth]{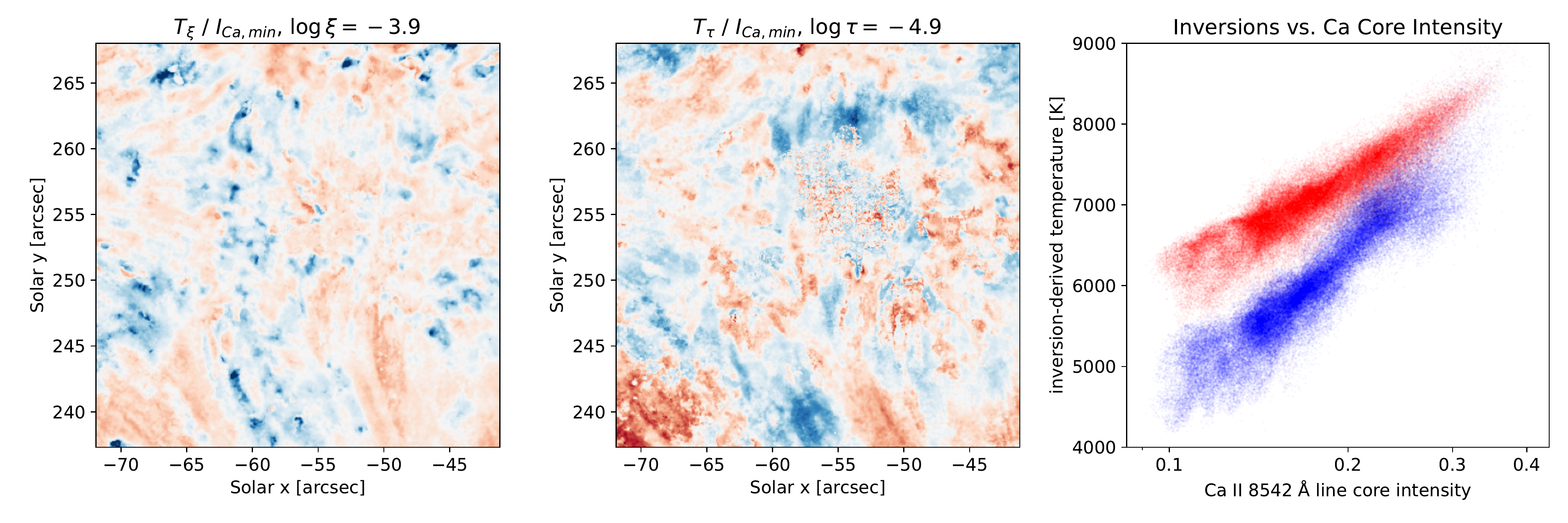}
    \caption{
    Maps of the ratios between the logarithm of the minimum intensity in the \ion{Ca}{2} 8542 \AA\ line core and inversion-derived temperatures, using the two different height scales, at heights of $\log \xi = -3.9$ \textit{(left)} and $\log \tau = -4.9$ \textit{(middle)}. Displayed values are scaled to $\pm 20\%$ of the mean ratio for each case.
    \textbf{Right:} scatter plots of inversion-derived temperatures, using the $\xi$ (red) or $\tau$ (blue) height scales, versus the \ion{Ca}{2} 8542 \AA\ line core minimum intensity, plotted on a logarithmic scale.}
    \label{fig:temp_ca_ratio}
\end{figure}

Further, the recovered temperature profiles from the optical depth inversions showed significant oscillatory behavior in the upper photosphere, most notably in the interval $-3 \le \log \tau \le -1$).
We also note that the inversions in optical depth tended to show a falling temperature in the chromosphere above the nominal optical-depth-unity height for the \ion{Ca}{2} 8542 \AA\ line core ($\log \tau \sim -4.8$), with some cases of anomalously low temperatures below 5000 K, while in the column mass inversions, the temperature profiles continued to rise above those heights.
In addition, the column mass inversions produce temperature structures that are better correlated with the \ion{Ca}{2} 8542 \AA\ line core intensity (Figure \ref{fig:temp_ca_ratio}), with outliers primarily consisting of fibril structures and localized cold spots not apparent in the line core minimum image. 
The minimum intensity of this line is known to be sensitive to the temperatures of the middle chromosphere \citep[see e.g.][]{2009A&A...503..577C, 2016MNRAS.459.3363Q}, so we can use it as a reasonable proxy to test our inversion results at the typical chromospheric heights of formation of the line core. 

\begin{figure}
    \centering
    \includegraphics[width=0.8\textwidth]{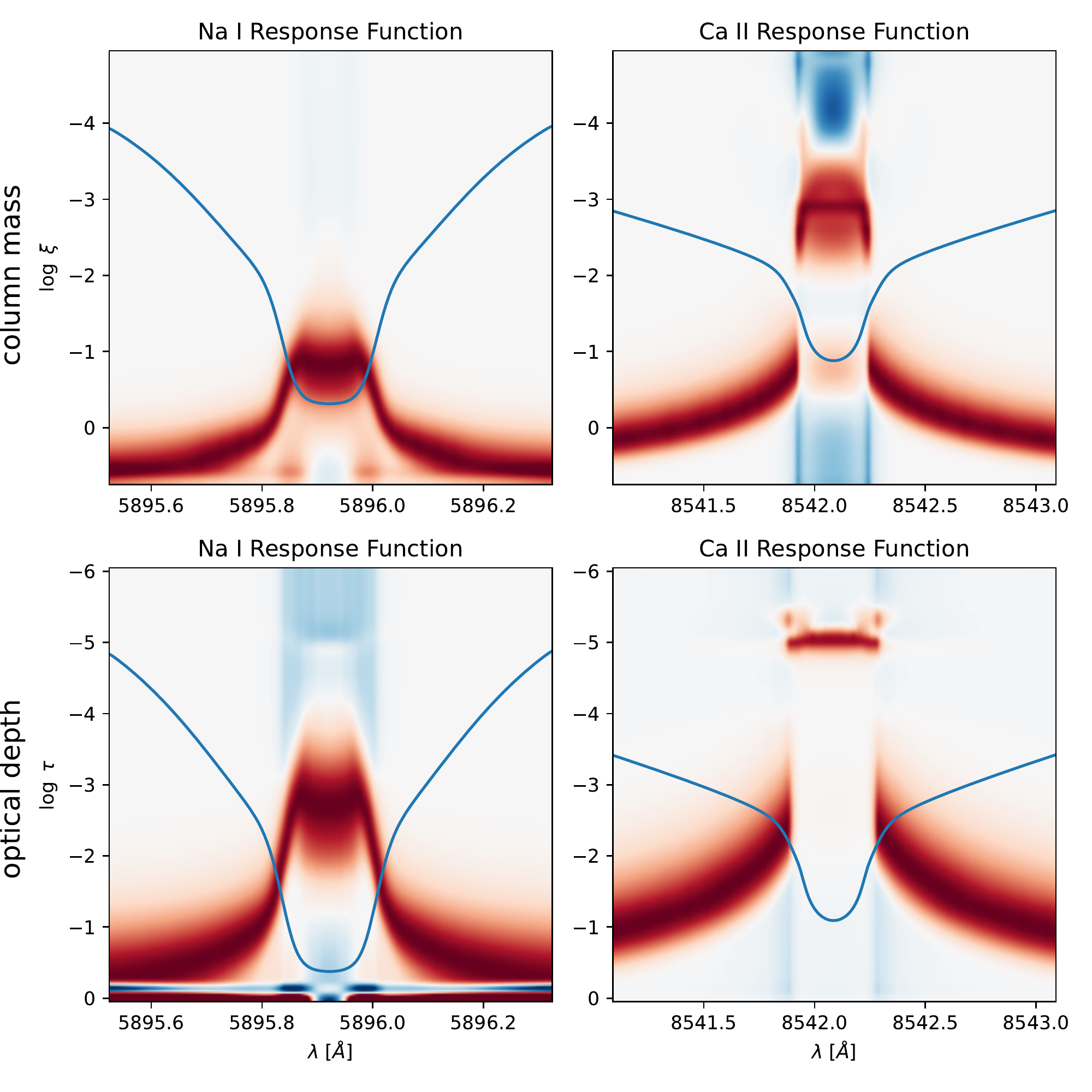}
    \caption{Response functions of the \ion{Ca}{2} 8542 \AA\ and \ion{Na}{1} 5896 \AA\ lines, computed for the FALC model in $\log\ \xi$ \textit{(top)} and $\log\ \tau_{500}$ \textit{(bottom)}. 
    The blue lines are the synthesized spectral profiles for each method.
    Note that the heights of the responses calculated in this standard model atmosphere are expected to differ from the heights inferred from the models derived through the inversion process.}
    \label{fig:falc_rf}
\end{figure}

To further illustrate the differences between these two alternative height scales, we computed the response functions in both height scales for the FALC model \citep[][]{1999ApJ...518..480F}, following the RF calculation methods from \citet[][]{2016MNRAS.459.3363Q}. These response functions, shown in Figure \ref{fig:falc_rf}, clearly show several key features discussed above: the photosphere is better sampled in optical depth, whereas the chromosphere has a greater extent in column mass; the sodium core is likewise more sensitive to temperatures over a broader range of optical depths; and the calcium line core response, which in $\log\ \tau$ is localized to a range of $\sim$ 0.2 dex, extends in $\log\ \xi$ over almost a full scale height. Furthermore, in contrast to the RFs computed by \citet[][]{2016MNRAS.459.3363Q}, the calcium line shows almost no sensitivity to temperature over the $\sim$ 1 dex range of the temperature minimum, corresponding roughly to $-4.5 \le \log\ \tau \le -3.5$ and $-2 \le \log\ \xi \le -1$.

Given our interest in studying the thermodynamic conditions in the temperature minimum and chromosphere, we chose to proceed with the analysis of the inversions done in column mass scale, as we consider them more reliable in this regime.

\subsection{Electron Density Model}
\label{sec:electrons}

\begin{figure}
    \centering
    \includegraphics[width=0.85\textwidth]{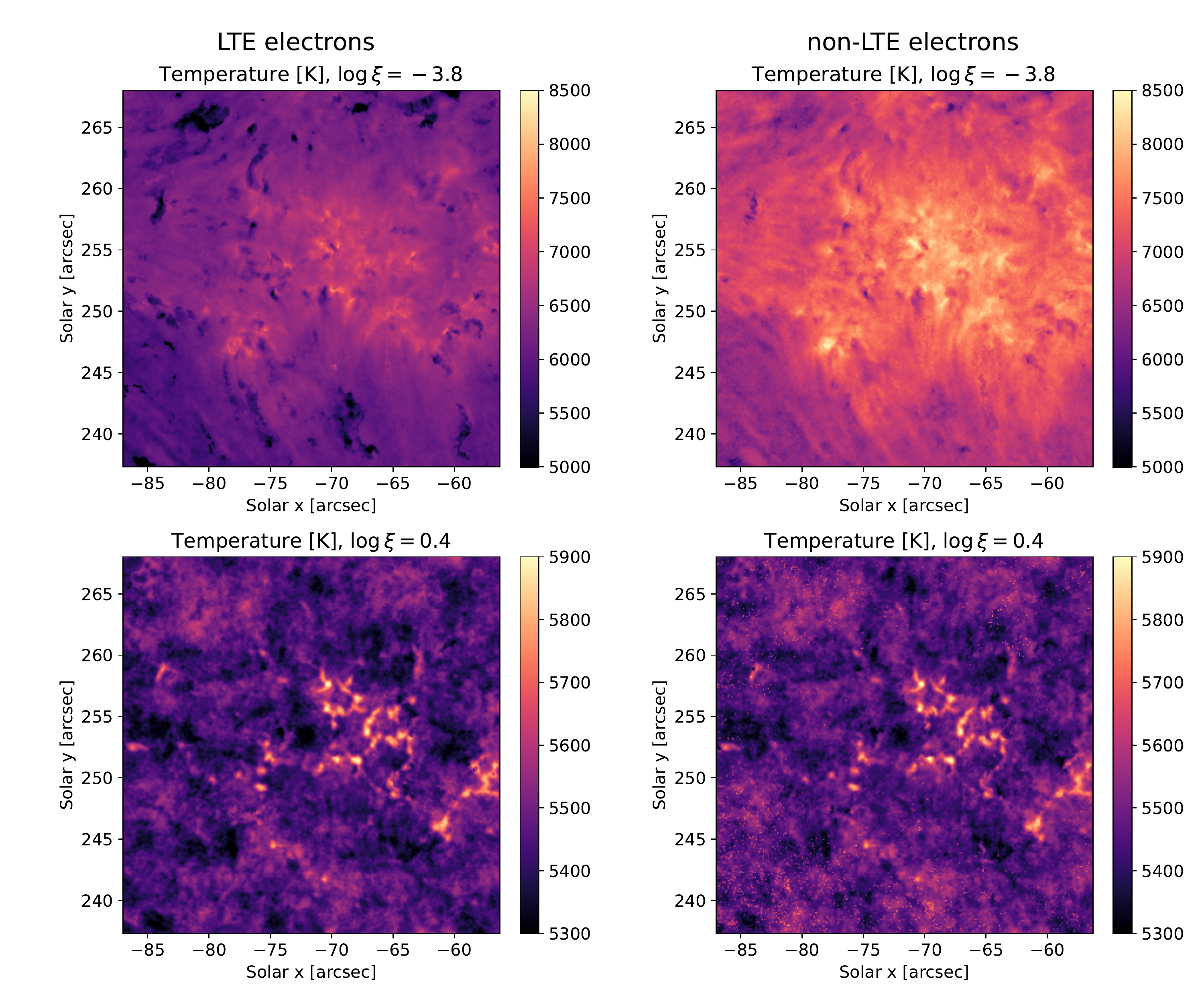}
    \caption{Temperature maps inferred from inversions of IBIS data for inversions without \textit{(left)} and with \textit{(right)} treatment of non-LTE hydrogen ionization. \textbf{Top row:} temperatures in low chromosphere ($\log \xi = -3.8$). \textbf{Bottom row:} temperatures in mid photosphere ($\log \xi = -0.4$).}
    \label{fig:nlte_maps}
\end{figure}

\begin{figure}
    \centering
    \includegraphics[width=0.85\textwidth]{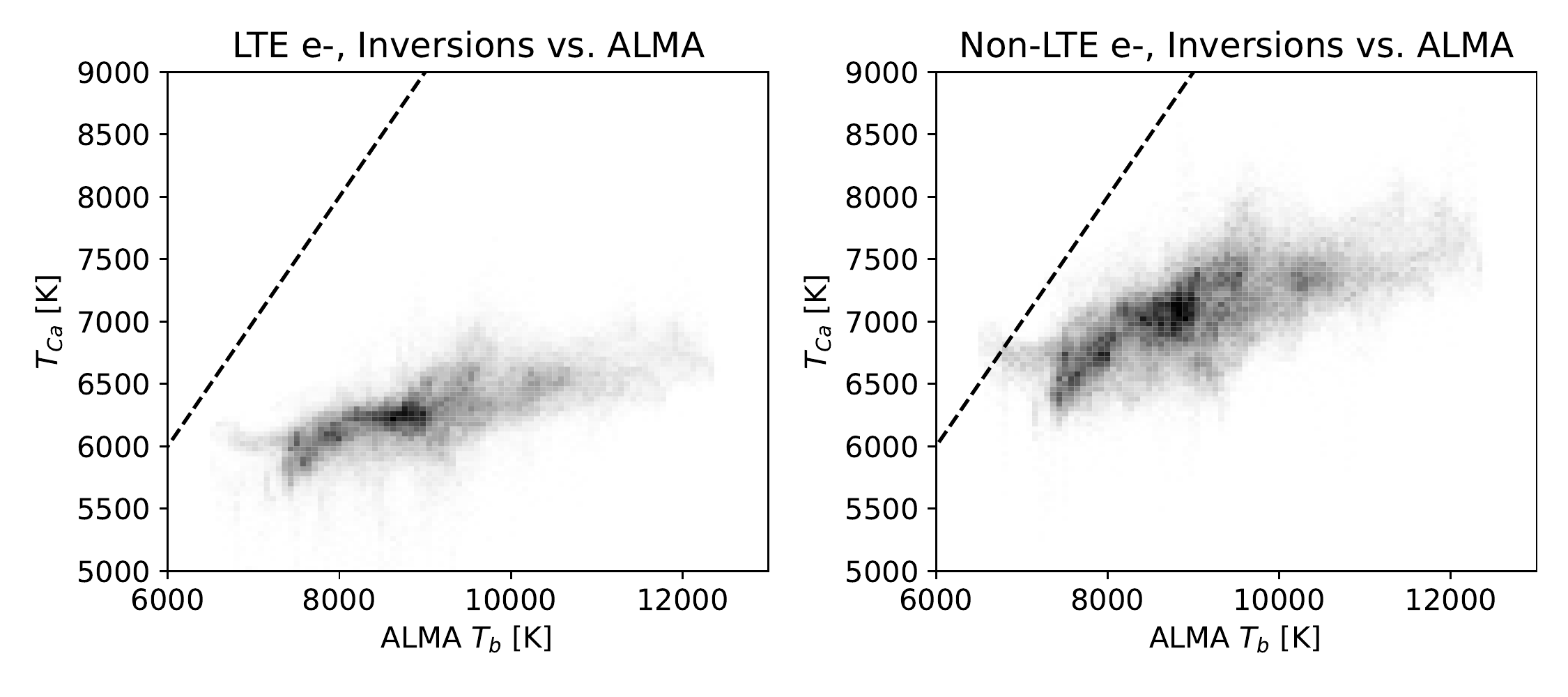}
    \caption{Relation between calcium line core temperatures and ALMA temperatures, without \textit{(left)} and with \textit{(right)} non-LTE hydrogen ionization.}
    \label{fig:nlte_scatter}
\end{figure}

It may be essential to take non-LTE hydrogen ionization effects into account in order to properly model electron densities when attempting to accurately determine chromospheric temperatures.
In the tenuous environment of the chromosphere, radiation is able to escape upwards, leaving fewer photons available to ionize hydrogen and thereby reducing the electron density for a given temperature. 
Therefore, when considering the non-LTE hydrogen ionization, the local electron density in the chromosphere is lower.  
This then reduces the coupling between the \ion{Ca}{2} source function and the local temperature, which would then produce a lower intensity in the line core. 
To properly reproduce the observed line core intensities, the inversion process must therefore assume higher chromospheric temperatures. 
The authors of STiC modified the RH synthesis module such that it can treat hydrogen as an active non-LTE species and iteratively solve for the electron pressure (albeit with a significant increase in the computational time required for each inverted pixel).

We used the column mass inversions above, both with and without non-LTE hydrogen ionization, to quantify this effect.
Figures \ref{fig:nlte_maps} and \ref{fig:nlte_scatter} illustrate this effect, with non-LTE hydrogen ionization increasing the inferred chromospheric temperatures by $\sim 1000$ K.
While we do not expect the calcium and ALMA temperatures to be identical, the inferred temperatures from the calcium line should at least be close to those measured from ALMA Band 6, given their similar expected heights of formation.
Inclusion of these physics primarily alters electron densities in the chromosphere, though significant departures from LTE electron densities can be observed as far down as the upper photosphere and temperature minimum, around $\log \xi = -2$ ($\log \tau = -4.1$).
Unfortunately, including this correction to the electron densities introduces significant numerical instability, which manifests as both salt-and-pepper "inversion noise" and occasional failures of individual profiles to converge.
We mitigated this issue by inverting with two cycles and strong regularization, and the atmospheric profiles of any pixels that failed to converge ($\sim 1\% $) were replaced with the median of the surrounding pixels.

\subsection{Final inversion configurations}
\label{sec:inversion_settings}

The final configuration for all our inversions combining the \ion{Ca}{2} and \ion{Na}{1} lines used [21, 11, 4] nodes in $[T, v_{\rm los}, v_{\rm turb}]$, equally spaced in column mass between $-5.1 \leq \log \xi \leq 0.9$, with regularization such that the penalty term in the merit function is typically of order unity.
We used spectral scans taken simultaneously with the ALMA data at 15:57 (series 1554) and 17:36 UT (series 1725).
For the series 1554 data, a simultaneous white-light continuum image is included to compensate for the poor sampling in the wings of the \ion{Na}{1} line.
The ALMA data, when included, are weighted similarly to the spectral data.

We performed the inversions of these datasets two times, once using only the IBIS spectral scans, and once including the cotemporal ALMA brightness temperature maps as inversion inputs, Band 6 in the case of the series 1554 scans and Band 3 with the series 1725 scans.

In addition to the observed ALMA data, we also took the opportunity to build on the work of \citet{2019ApJ...881...99M}, specifically in the nearly linear relationship between the H-$\alpha$ line width and the Band 3 brightness temperature.
As all the data taken with IBIS also included H-$\alpha$ line scans, we were able to use a snapshot of the line width as a pseudo-band 3 image for the series 1554 data.
The line-width snapshot was blurred to the resolution of ALMA Band 3 and converted to an approximate 3-mm brightness temperature using the relation in their work.
We tested the effects of including this proxy data in the inversions, both with and without the real observed Band 6 data, in Section \ref{sec:both_ALMA_inversions}.

\section{Comparisons between spectroscopic inversions and ALMA}
\label{Ch:ALMA_comparisons}

One primary aim of this work is to study the agreement between the temperatures inferred from spectroscopic inversions and the brightness temperatures observed with ALMA.
With two largely independent measures of temperature, each with a different mechanism of formation, we have the means to not only test the fidelity of the temperatures recovered in the inversion process, but also to test our understanding of the formation of the millimeter continuum.
This is particularly interesting because we are using the \ion{Na}{1} D and \ion{Ca}{2} 8542 \AA\ lines, whose heights of formation overlap significantly with the regions of the atmosphere contributing to the millimeter emission in the observed bands.
The \ion{Ca}{2} 8542 \AA\ line in particular is widely used in studies of the chromosphere, often employing similar kinds of inversion methods \citep[see e.g.][]{2018A&A...620A.124D, 2019A&A...623A..74D, 2018A&A...617A..24M}.
We can also address the question of whether including observations in Band 3 or Band 6 (or both) in the inversion process might provide more realistic constraints when using these spectral lines.

\subsection{ALMA observations and inversion-derived temperatures}
\label{sec:Band_3_vs_inversions}

\begin{figure}
    \centering
    \includegraphics[width=\textwidth]{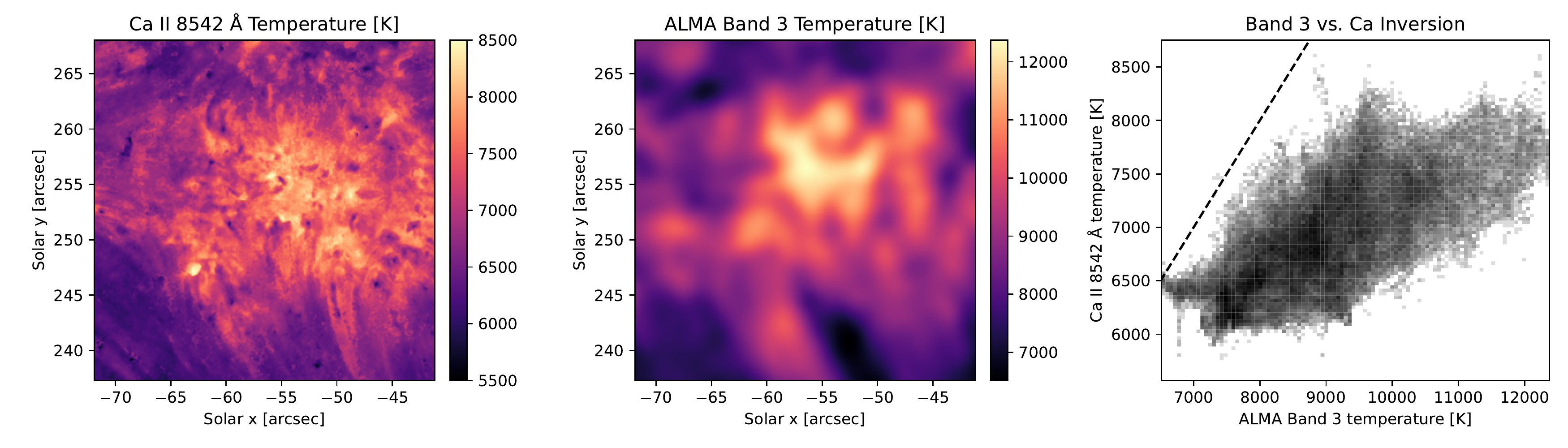}
    \caption{Comparison of ALMA Band 3 observations with inversion results for series 1725 IBIS observations. \textbf{Left:} Temperatures inferred from inversions at $\log \xi = -3.8$ (corresponding roughly to the \ion{Ca}{2} 8542 \AA\ line core). \textbf{Middle:} Observed ALMA Band 3 brightness temperatures. \textbf{Right:} Density plot of temperatures derived from spectroscopic inversions and ALMA observations, with the 1:1 correlation indicated by the dashed line.}
    \label{fig:inv_vs_alma_band3}
\end{figure}

\begin{figure}
    \centering
    \includegraphics[width=\textwidth]{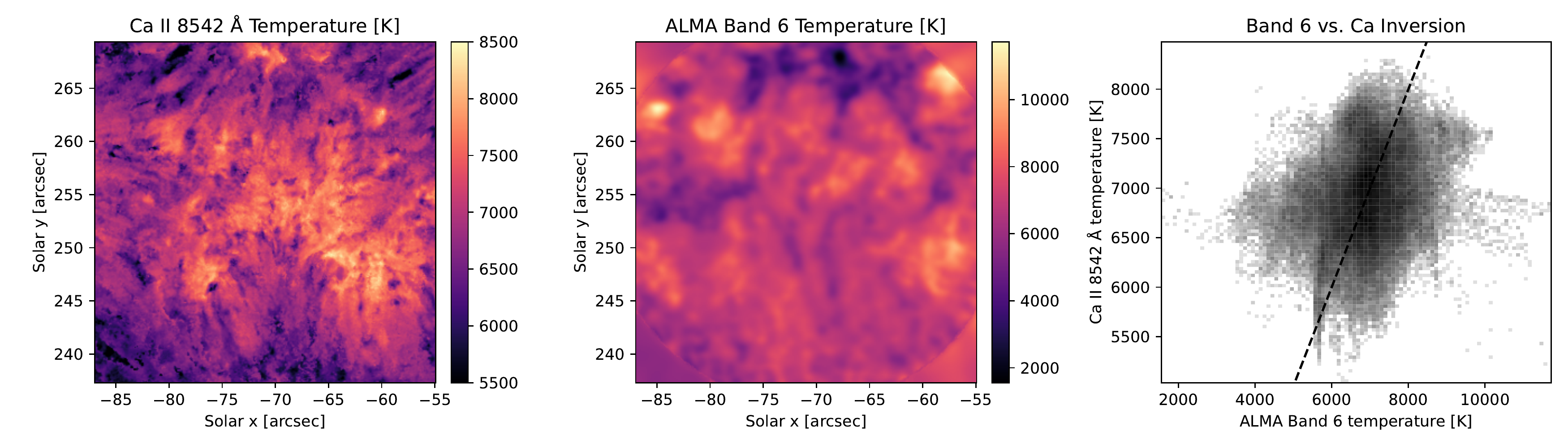}
    \caption{Comparison of ALMA Band 6 observations with inversion results for series 1554 IBIS observations. \textbf{Left:} Temperatures inferred from inversions at $\log \xi = -3.8$ (\ion{Ca}{2} 8542 \AA\ line core). \textbf{Middle:} Observed ALMA Band 6 brightness temperatures. \textbf{Right:} Density plot of temperatures derived from spectroscopic and ALMA observations, with a 1:1 correlation indicated by the dashed line.}
    \label{fig:inv_vs_alma_band6}
\end{figure}

We begin by comparing the ALMA Band 3 image, taken at 17:36 UT, with the inversion-derived temperatures from the cotemporal spectral line observations.
Figure \ref{fig:inv_vs_alma_band3} shows the comparison of the inversion-derived temperature at $\log\xi=-3.8$ (roughly corresponding to $\log\tau=-5.0$) and the ALMA band 3 brightness temperature. 
This specific height was chosen because the inverted temperatures show the best correlation with the core intensity (depth) of the \ion{Ca}{2} 8542 \AA\ line profile, indicating this is the column mass layer corresponding, on average, to the peak of the response function \citep[e.g.][]{2016MNRAS.459.3363Q, 2018A&A...620A.124D}. 
While the maps are similar on large spatial scales, with bright/hot plage in the center of the field and hotter fibrils extending outward into the internetwork region below, they differ significantly on smaller scales.
Given the lower spatial resolution of the ALMA Band 3 data (approximately 2\arcsec), it is not clear whether the small scale structures seen in the Ca-II-derived maps would similarly be present in the Band 3 continuum if higher resolution could be achieved. 

Furthermore, it can be seen that the ALMA temperatures are systematically higher than the temperatures inferred from inversions.
This is not unexpected, as previous works \citep[e.g.][]{2015A&A...575A..15L,2016SSRv..200....1W} show that the Band 3 (3.0 mm) radiation is typically expected to form at slightly higher chromospheric layers than the \ion{Ca}{2} core, in a region where we expect the transition region temperature rise to have begun. We note that the difference in absolute temperatures is significantly larger than the expected errors in the absolute calibration \citep{2017SoPh..292...88W} or the expected offsets between the different sub-bands \citep{2020A&A...635A..71W}.
Nonetheless, the large-scale structures are fairly similar, with Pearson correlation coefficient $R = 0.69$.

We find that the correlation coefficient between the inverted temperatures and ALMA brightness temperature maps falls off rapidly at greater heights, for $\log\xi<-3.8$, dropping nearly linearly from the peak to 0.2 at the upper bound of our inverted atmospheres ($\log\xi=-5.1$).
This indicates that caution should be taken when interpreting the inverted atmospheres as they become unreliable above the peak of the response function of the highest-forming line. This is expected, and here we confirm it by using the independent temperature measurement from ALMA.

In a similar manner, we compare the ALMA Band 6 image, taken at 15:57 UT, with the inversion-derived temperatures.
Figure \ref{fig:inv_vs_alma_band6} shows the comparison between inverted temperatures at $\log \xi = -3.8$ and the ALMA band 6 brightness temperatures.
In this case, the temperatures for these two diagnostics
are much closer to each other in terms of mean value (7000 K and 6900 K, respectively), but the correlation between them is significantly weaker (only 0.22), as seen in the large cloud of points in the scatterplot on the right side of the figure.
There are no other values of $\xi$ with a significantly improved correlation with the inversion temperature. 
While our average Band 6 temperature is similar to the value of 7000 K found by \citet{2019A&A...622A.150J} for plage regions surrounding a sunspot, those authors found a much higher correlation coefficient with the radiation temperature of the upper chromospheric \ion{C}{2} line (0.83). Similarly, \citet{2017ApJ...845L..19B} found temperatures of $\sim$7300 K for their definition of the plage areas in this same dataset, and a good correlation with the \ion{Mg}{2} h line radiation temperature (0.71). 
However, we note that the correlations in these two studies were based on just a restricted portion of the field of view which tended to select regions of higher Band 6 temperatures, whereas our correlation is calculated over our full field and spanning a broader range of temperatures. A similar result was was found by \citet{2021ApJ...906...83C} who noted the sensitivity of these measured correlations to the definition of the plage mask.
As discussed in Section \ref{sec:ALMA6_and_spectra_inversions}, the regions of high Band 6 temperatures are indeed well reflected in the inversion-derived temperatures maps for the upper atmosphere, but the correspondence breaks down when the full span of observed temperatures is considered. The Band 6 contribution functions can vary over a significant range of physical heights, based on the local temperatures, resulting in a lack of overall correspondence between the brightness temperatures and the temperature structures at a single height.

Previous radiative transfer calculations using MHD simulations have shown that the heights of formation of the Band 6 (1.2 mm) emission may spread, even for the same column, from the temperature minimum region to  heights above that of the \ion{Ca}{2} 8542 \AA\ line core \citep{2015A&A...575A..15L,2018A&A...620A.124D}.
However, recent calculations by \citep{2020ApJ...891L...8M}, including the effects of nonequilibrium and time-dependent ionization of hydrogen and helium, found that the Band 6 radiation should form at a narrower range of heights very similar to that of the Band 3 radiation. Yet, in our case, while the ALMA temperature maps in Figure \ref{fig:inv_vs_alma_band3} and Figure \ref{fig:inv_vs_alma_band6} show some similar large-scale structuring following the plage magnetic fields, the details on smaller scales seems to be quite different and not entirely consistent with the idea that they are formed at nearly coincident heights in the atmosphere.
Some of these differences may be due to the evolution of the chromospheric fine structure in the 100 minutes between the times when these two observations were acquired. However, we examined the magnitude of the changes seen in either band over the full one-hour duration of their respective observing blocks, and we find that the magnitude of the evolutionary changes over these timescales are smaller than the differences seen between the two bands.
The distinct differences between the inversion correlations for these two bands in our dataset indicate there may be more significant variations in the formation of these two continua then expected.

The ALMA Band 6 temperature map shows some very cold pixels ($\lesssim$3000 Kelvin), especially at the top of the field of view, that do not exist in the spectroscopic inversions (STiC enforces 3100 Kelvin as the lowest possible temperature because chemical equilibrium solvers cannot converge at lower temperatures).
These cold regions are not present in the maps of inversion-derived temperatures at chromospheric heights, suggesting that the classical spectral lines in the visible and near-infrared might not be highly sensitive to such cold pockets in the atmosphere.
This is supported by \citet[][]{2012A&A...543A..34D}, who found a similar result for inversions of synthetic \ion{Ca}{2} data, attributing this insensitivity to 3-D scattering effects.
Indications of similar cold spots have been identified using observations of the CO 4.7 $\mu m$ line \citep[][]{2022ApJ...930...87S}, though it remains unclear at which heights the CO is forming.

\subsection{Simultaneous inversions of spectroscopic data and ALMA Band 3 data}
\label{sec:ALMA3_and_spectra_inversions}

\begin{figure}
    \centering
    \includegraphics[width=\textwidth]{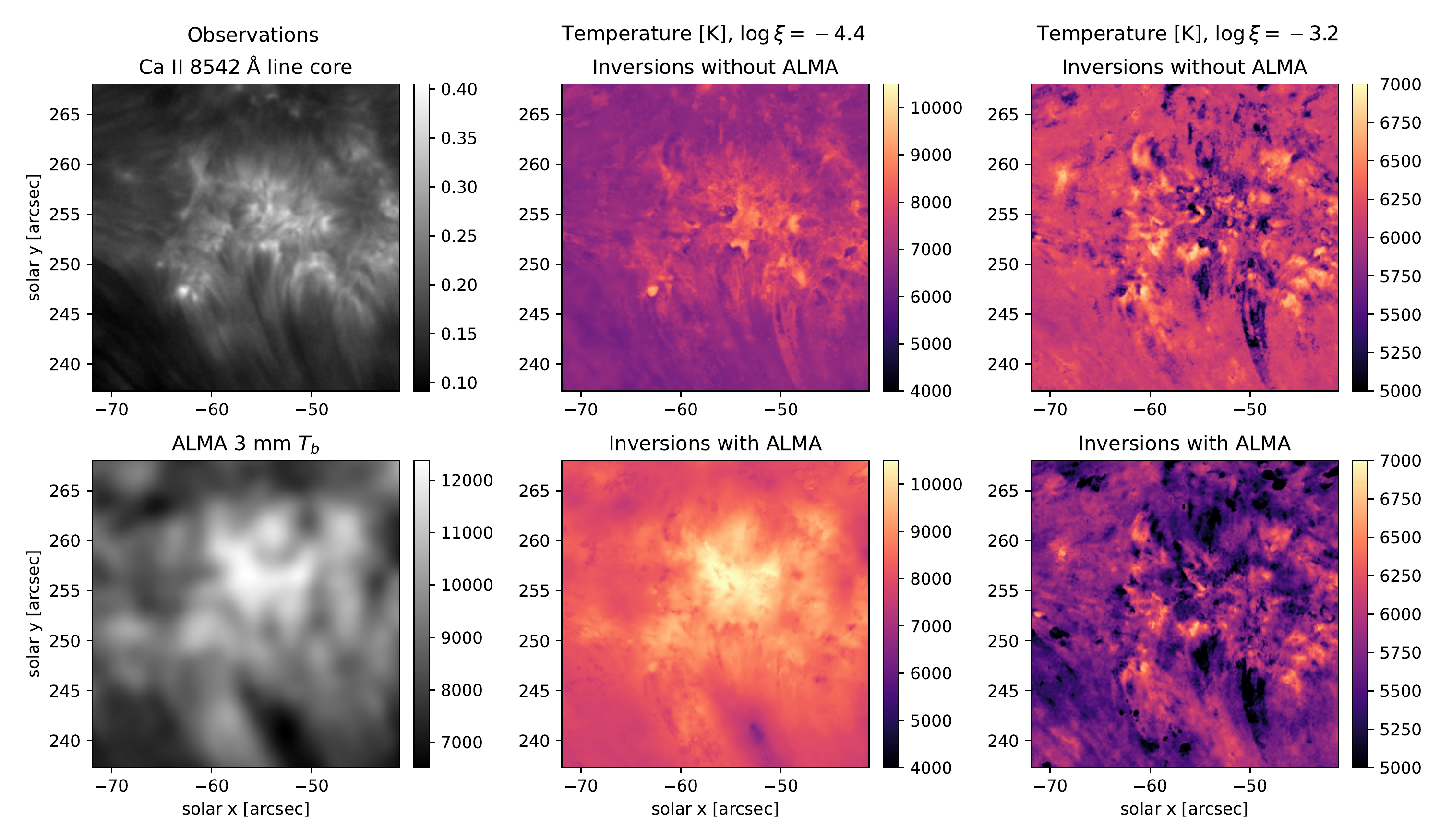}
    \caption{Effects of including Band 3 $T_b$ in the inversion of the series 1725 data. \textbf{Top:} inversion-derived temperatures using only IBIS spectroscopic data. \textbf{Bottom:} inversion-derived temperatures using both IBIS and ALMA Band 3 data. Temperature maps are shown for two heights, located  at 0.6 dex above \textit{($\log \xi = -4.4$, center)} and below \textit{($\log \xi = -3.2$ right)} the \ion{Ca}{2} 8542 \AA\ line core.}
    \label{fig:nlte_alma3_comp}
\end{figure}

To test whether the inclusion of the millimeter continuum as an additional input in our inversion process leads to more reliable atmospheric profiles, we inverted the series 1725 spectroscopic data together with cotemporal ALMA Band 3 data.
The Band 3 continuum is expected to be formed in the upper chromosphere, therefore it is expected to act mostly as a constraint on the temperatures in the upper chromosphere, with less significant effects on the temperatures in lower regions to which the \ion{Ca}{2} 8542 \AA\ line is sensitive.
This was explored by \citep[][]{2018A&A...620A.124D}, who compared inversions of synthetic \ion{Ca}{2} 8542 \AA\ (and \ion{Fe}{1} 6302 \AA) profiles with and without the ALMA Band 6 and Band 3 continua (their inversion setup i3 and i4).
They found that without the Band 3 continuum, the derived temperatures at regions above the height of formation of the \ion{Ca}{2} 8542 \AA\ line are poorly constrained and unrealistic.
We find similar behavior in our inversions, as seen in Figure \ref{fig:nlte_alma3_comp}, where the central column shows the inversion-derived temperatures at $\log \xi = -4.6$ (roughly corresponding to the $\log \tau = -5.6$ shown in \citet[][]{2018A&A...620A.124D}). When only the spectral lines are inverted (upper row), the temperature profiles in the higher layers do not recover any significant temperature increase in the upper chromosphere.
However, when the ALMA Band 3 temperatures are included, the inferred temperatures show the expected increase of temperatures in the upper atmosphere, ranging from approximately 7000--12000 K.

As expected, the changes in the temperatures at lower heights, around the height of formation of the \ion{Ca}{2} 8542 \AA\ line core, are less dramatic, but it does appear to increase the dynamic range of temperatures in these regions.
The right column of Figure \ref{fig:nlte_alma3_comp} shows the temperatures at $\log \xi = -3.2$ (roughly $\log \tau = -4.4$) when inversions were done without and with the Band 3 brightness temperature.
When the millimeter continuum is included, the average temperatures at this height appears to be lowered, from T $\sim$ 6500 K to 5800 K. 
Some of the smaller-scale structuring of the higher temperatures in the spectral-line-only inversions appear to be replicated when Band 3 is included. 
These inversions also introduce regions of significantly lower temperature at this height, that are not seen in the inversions without the Band 3 (e.g. x=-54, y=261), sometimes corresponding to regions of higher temperatures in the Band 3 temperatures. There is also a region centered on x=-51, y=240, where the lower temperatures seen at $\log \xi = -4.4$ correspond to an increased temperature at $\log \xi = -3.2$. 
Some of the changes seen at this lower height may be unrealistic or artifacts of the inversion technique, raising the question of whether the inclusion of the millimeter continuum observations may have mixed (height-dependent) effects on the reliability of the recovered atmospheres.
A similar effect was seen in \citet[][]{2018A&A...620A.124D} with synthetic inversions including both the Band 6 and Band 3 continua together with the \ion{Ca}{2} 8542 \AA\ line, but we see evidence of similar issues even with only the higher-forming Band 3.

\subsection{Simultaneous inversions of spectroscopic data and ALMA Band 6 data}
\label{sec:ALMA6_and_spectra_inversions}

\begin{figure}
    \centering
    \includegraphics[width=\textwidth]{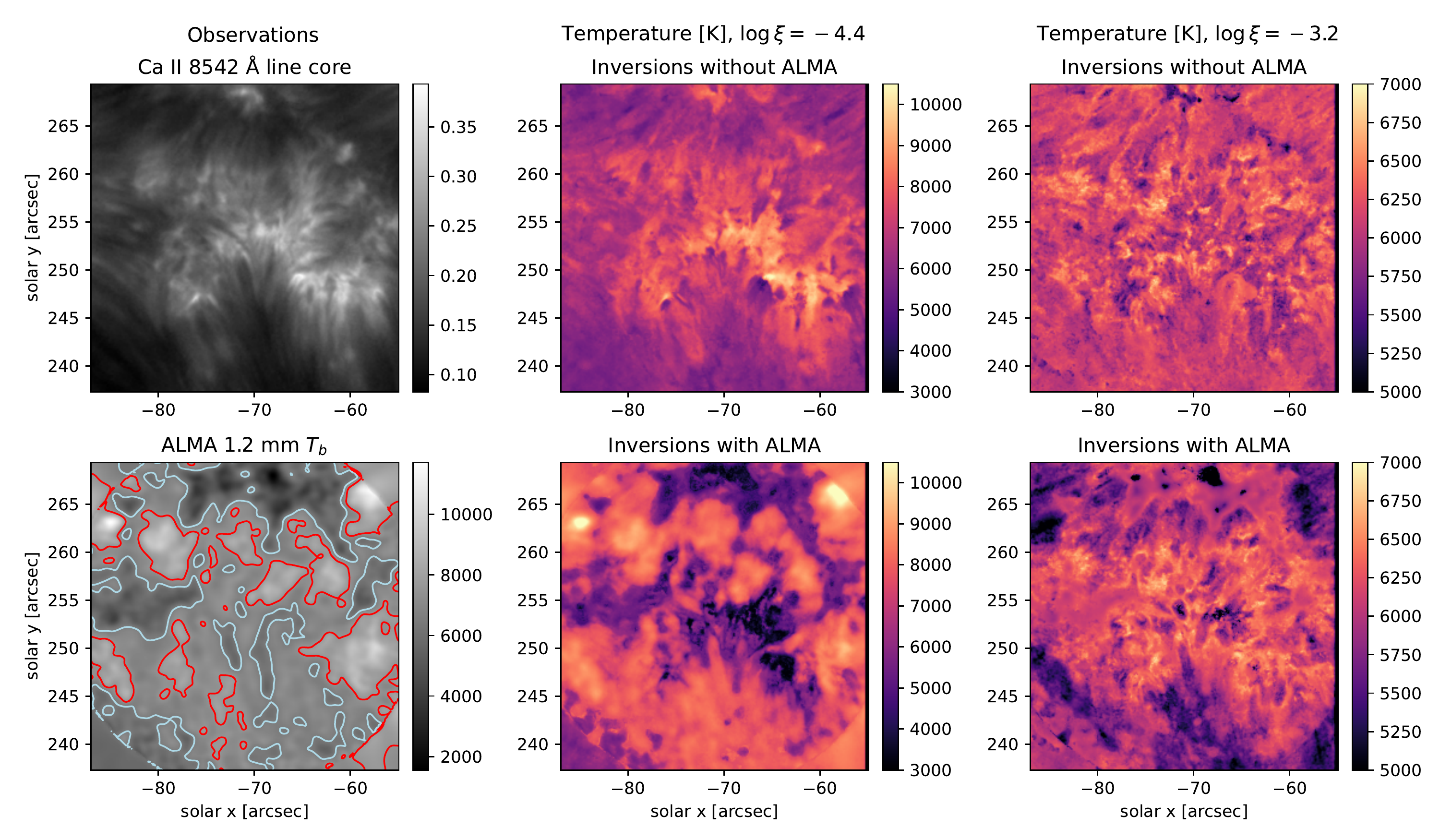}
    \caption{Effects of including Band 6 $T_b$ in the inversion of the series 1554 data. \textbf{Top:} inversion-derived temperatures using only IBIS spectroscopic data. \textbf{Bottom:} inversion-derived temperatures using both IBIS and ALMA Band 6 data. Temperature maps are shown for two heights, located 0.6 dex above \textit{(($\log \xi = -4.4$, center)} and below \textit{(($\log \xi = -3.2$, right)} the \ion{Ca}{2} 8542 \AA\ line core. Contours in lower left plot indicate "hot" ($>$ 7500 K, red) and "cool" ($<$ 6500 K, blue) regions for which response functions were computed.}
    \label{fig:nlte_alma6_comp}
\end{figure}

\begin{figure}
    \centering
    \includegraphics[width=\textwidth]{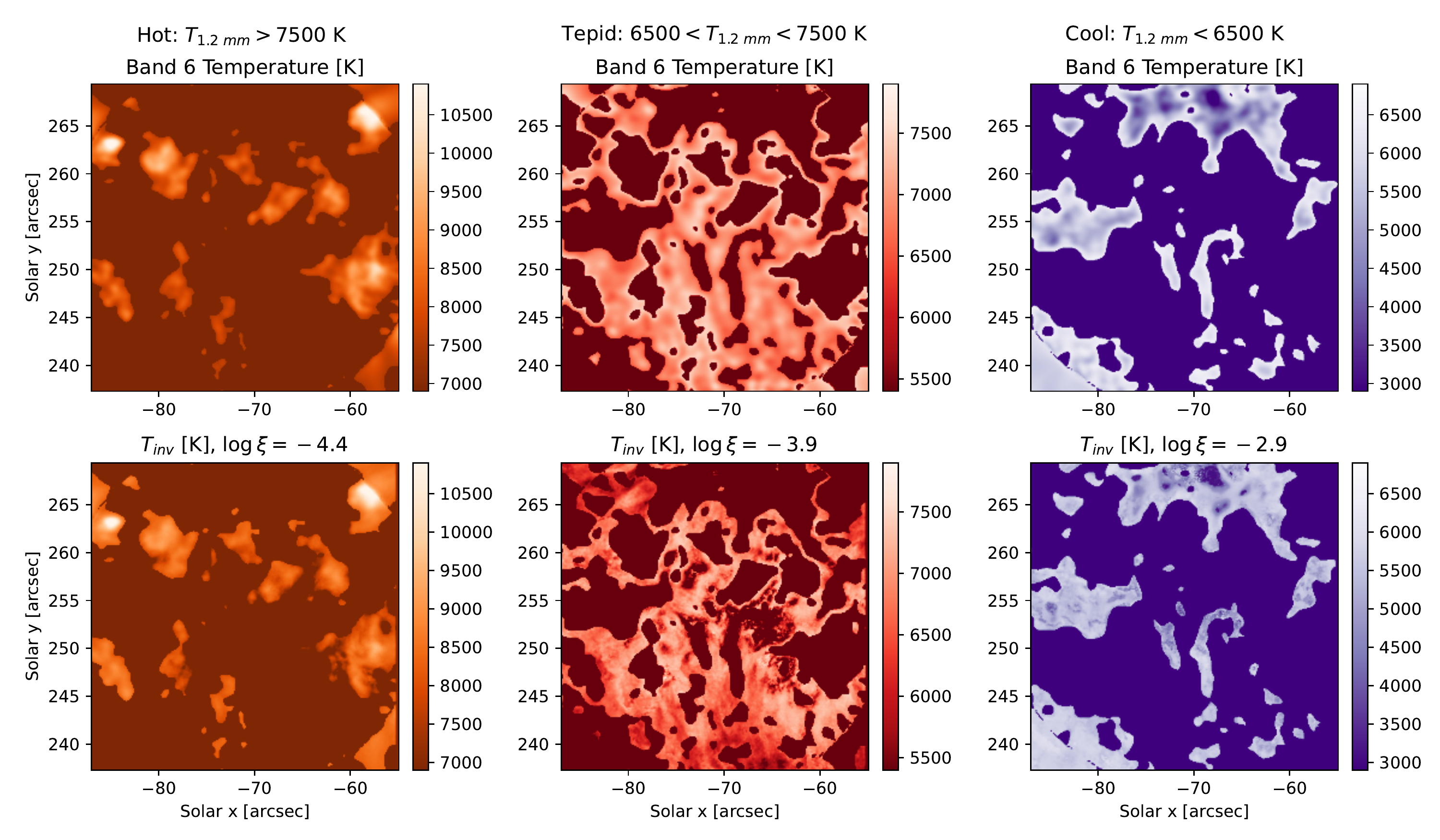}
    \caption{Comparison of ALMA observations with joint IBIS and ALMA inversion results for series 1554 observations. \textbf{Top:} ALMA Band 6 $T_b$, segmented into hot ($T_b > 7900$ K), tepid or intermediate ($5900 \leq T_b \leq 7900$ K), and cold ($T_b < 5900$ K) regions. \textbf{Bottom:} temperatures inferred from joint inversions, sampled at heights of best correspondence with ALMA data.}
    \label{fig:ca_vs_band6}
\end{figure}

\begin{figure}
    \centering
    \includegraphics[width=\textwidth]{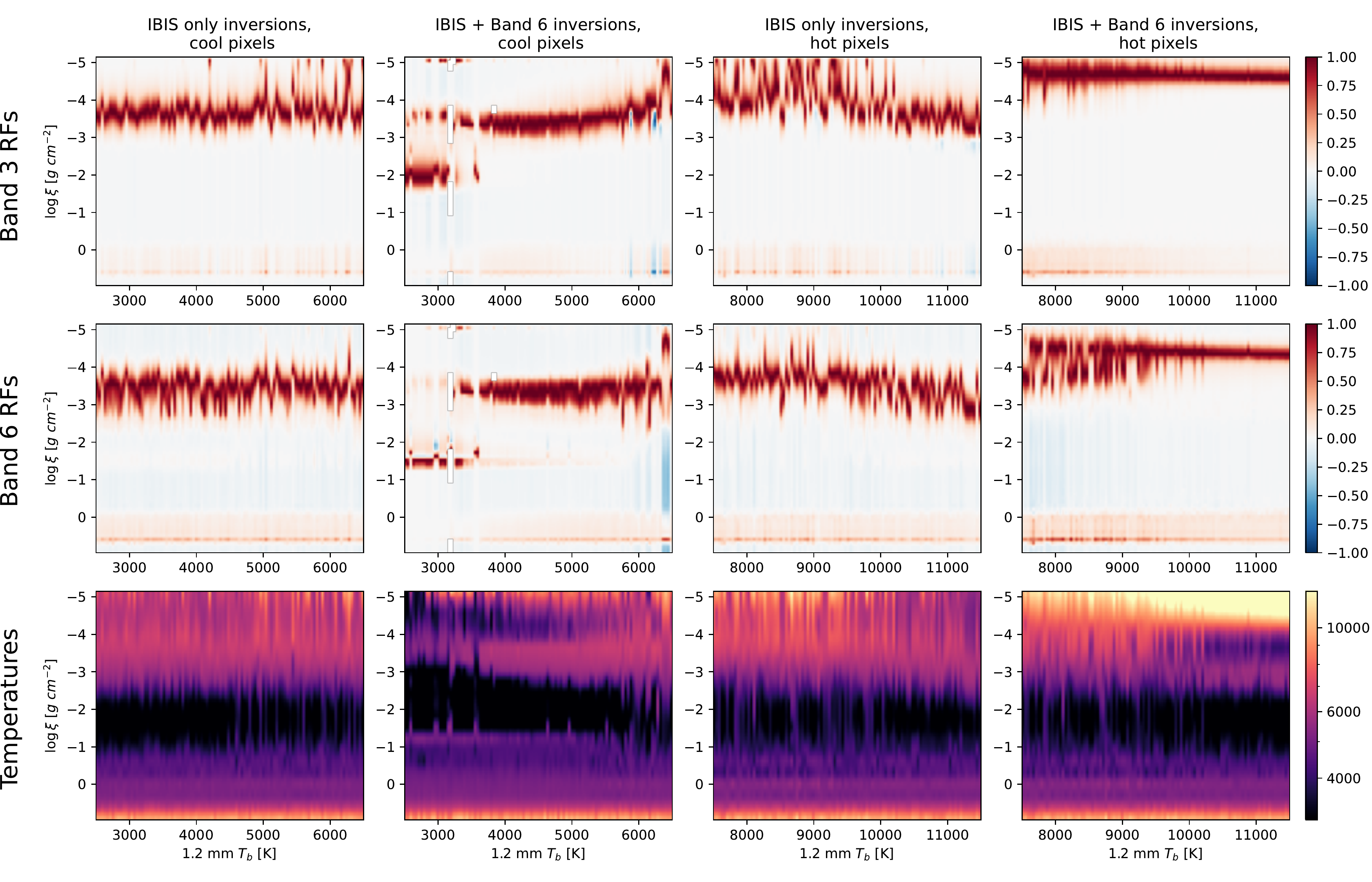}
    \caption{Response functions of Band 3 \textit{(top)} and Band 6 \textit{(middle)} continuum, computed for the inversions of spectroscopic data without \textit{(left)} and with \textit{(right)} ALMA Band 6 as input. 
    The RFs are plotted in order, sorted left-to-right by their Band 6 $T_b$. \textbf{Bottom:} temperature stratifications from which response functions were computed, sorted in the same way to facilitate a direct comparison.}
  \label{fig:band_6_RF}
\end{figure}

We next explore the effects of including ALMA Band 6 observations in the inversions (instead of Band 3), using the series 1554 IBIS spectral scans together with cotemporal ALMA Band 6 data. 
This shorter wavelength is expected to typically form lower in the chromosphere, overlapping to a greater extent with the formation region for the \ion{Ca}{2} 8542 \AA\ line and perhaps \ion{Na}{1} 5896 \AA\ as well \citep[][]{2015A&A...575A..15L}, though \citet[][]{2021ApJ...906...83C} point out that in some dynamic phenomena the Band 6 emission might come from much higher in the atmosphere. 
Therefore, we expect this additional diagnostic to provide stronger, or more consistent, constraints for the inversion process. 

Examining the results shown in Figure \ref{fig:nlte_alma6_comp}, we see that the most significant differences in the recovered temperatures are at the greater heights, $\log \xi = -4.4$ (central column).
In many cases, these inversions produce seemingly unphysical atmospheres, introducing extended regions of extremely cold ($\sim 4000$ K) gas in the upper-chromosphere ($\log \xi > -4.0$), above the warmer ($\sim 8000$ K) gas at $\log \xi = -3.8$ needed to match the \ion{Ca}{2} 8542 \AA\ line core. This occurs both in cases where the Band 6 temperatures are low (such as the extended cool regions at the top or at the left of field of view) and where the temperatures are close to the average (for example, the extended $\log \xi = -4.4$ cold region at x=-65 and 245\,$<$\,y\,$<$\,255).
We note that the temperatures observed in Band 3, expected to be formed roughly at this height, doesn't show such low values, nor are these low temperatures consistent with those recovered from the inversions shown in Sec. \ref{sec:ALMA3_and_spectra_inversions}.
The inversion-derived temperatures at $\log \xi = -3.2$ shown in the right column of Figure \ref{fig:nlte_alma6_comp} show less significant differences compared to the spectral-line-only inversions. As with the inversions including Band 3, there does appear to be an puzzling inverse correlation in some locations, with lower temperatures at this height in correspondence to higher temperatures in the upper layers (e.g. x=-85, y=263 or -86\,$<$\,x\,$<$\,81 and 243\,$<$\,y\,$<$\,250). 

We note that STiC has a hard lower limit on the allowable temperatures of 3100 K.
Even if the inversion code inferred that the cold Band 6 temperatures were coming from lower in the atmosphere, it might be difficult for it to properly fit atmosphere to some of the very low Band 6 temperatures (T $<$ 4000 K). 
Looking back at Figure \ref{fig:xvt_hist}, we see that, even from spectroscopic-only inversions, a significant amount of pixels reach this hard-set minimum temperature around column masses $\log\,\xi =$ -1.0 to -2.0; however, this is more likely due to a lack of temperature sensitivity at these heights, combined with the tendency of the regularization to prefer smooth temperature profiles.

More problematically, there is no single atmospheric layer (or value of $\xi$) where the inverted temperatures on aggregate correlate well with the ALMA band 6 brightness temperatures.
This can be seen in Figure \ref{fig:ca_vs_band6}, where we have partitioned the Band 6 brightness temperature map into three regions based on the observed Band 6 brightness temperatures. We define ``hot'' ($T_b > 7900$ K), ``tepid'' ($5900 < T_b \leq 7900$ K), and ``cold'' ($T_b \leq 5900$ K) masks. We then applied those masks to the cubes of inversion-derived temperatures and for each case identified the column mass layer that had the best correlations with the Band 6 temperatures encompassed in that same mask. 
We found that the hottest regions seen in Band 6 ($T_b \gtrsim 7900$ K) correlated best with temperatures at $log\,\xi = -4.4$, which falls slightly above the \ion{Ca}{2} 8542 \AA\ line core (Pearson correlation coefficient, R = 0.4). The coldest regions in the brightness temperature maps instead seem to form lower down in the atmosphere, at $log\,\xi = -2.9$, just above the temperature minimum (R = 0.7). This indication of multi-height formation can be seen in the close visual similarity of the structures seen under the hot and cold masks in both the observed Band 6 brightness and the inverted temperatures. Interestingly, for the ``tepid'' regions in the middle column, while the best correlation is at $log\,\xi = -3.9$, there is less similarity between the structures in the two temperature maps (R = 0.1). This lack of correlation with a specific height in the atmosphere might indicate that the Band 6 $T_{b}$ for these intermediate temperatures are not generally representing a single or well-constrained region of the solar atmosphere, but might be mixing contributions from multiple heights.

One way to determine from which layers the mm-continuum originates is to analyze the response functions (RFs) for the mm-continuum (Figure \ref{fig:band_6_RF}).
To do this, we sorted the pixels in order of increasing $T_b$, then selected 100 pixels at regular intervals of 40 K, spanning from 2500--6500 K (labeled cool) and 7500--11500 K (labeled hot). Given the general level of uncertainty in the Band 6 temperature values, this gave a randomly distributed selection of points in these two different temperature regimes.
For each pixel selected, we extracted the 1-D atmospheric models from the inversions of the series 1554 data, both with and without the inclusion of Band 6, and computed the RFs for both 1.2 and 3.0 mm millimeter continuum.
The resultant RFs for both millimeter bands are shown in Figure \ref{fig:band_6_RF}, along with their respective temperature stratifications, with the columns comparing the two inversion schemes (with and without Band 6).

The most noticeable difference is the significant increase of the peak of the response functions ($\log\,\xi < -4.5$) in the hot pixels for both the 1.2 and 3.0 mm continuum when the observed Band 6 temperatures are included in the inversion process (rightmost column). 
Without Band 6 as an inversion input, only a sampling of the 3.0 mm RF's and few of the Band 6 RF's extend to heights of $\mathrm{log}\,\xi < -4.0$. The temperature profiles in the bottom row show that the inversions of the \ion{Ca}{2} 8542 \AA\ line only capture the expected strong rise in temperatures at the top of the chromosphere if also provided with additional information about the presence of high temperatures as observed in the ALMA Band 6. 
It is interesting that these IBIS and Band 6 inversions show, for hot pixels in the range 7500--9500 K, a generally extended hot chromosphere from $-2.8 > \log\,\xi > -5.0$, with the Band 6 RF's often extending over a large part of this range. Instead for the pixels with Band 6 temperatures above 9500 K the most common solution from the inversions indicates an initial temperature rise around $\log\,\xi = -3.0$, followed by a temperature drop at ${\log}\,\xi \sim -3.8$ (the approximate height of formation of the \ion{Ca}{2}  8542 \AA\ line), and then a very steep temperature rise above $\mathrm{\log}\,\xi = -4.2$. The Band 6 RFs in this case are very localized in height to this region of a steep temperature gradient.
While this might represent real differences in the chromospheric conditions, the significant bifurcation in the class of temperature profiles might indicate that instead these are simply different solutions from the inversion process when trying reconcile the emissions from the spectral lines and millimeter continuum. 

For the cool pixels in the left two columns, the biggest change in the RF's occur for the lowest temperature pixels ($T_b < 3500 K)$. For these cases, when including Band 6 in the inversion process, the peak of the RF's are moved very low in the atmosphere, to $\mathrm{log}\,\xi = -1.5$, the very heart of the temperature minimum in the atmosphere. While the inversions without Band 6 as an input reach similarly low temperatures in the middle atmosphere, when constrained by the presence of very low Band 6 $T_b$, solutions must be found that keep the upper atmosphere relatively cool and compressed in order to minimize the millimeter continuum opacity in the upper atmosphere. Otherwise, the 1.2 mm radiation would be formed higher up in the atmosphere, where the \ion{Ca}{2} 8542 \AA\ line places limits on the allowable minimum temperatures. Even with these atmospheric modifications, the inversions for these coldest pixels do not fully reproduce the extreme low temperatures seen in the Band 6 observations.
For many of the other pixels, the RF's show similar average heights $\mathrm{log}\,\xi \sim -3.5$ for both the 1.2 and 3.0 mm radiation. However, the RF's are more uniform and sharply peaked for the inversions that include the Band 6 inversions, primarily because the solutions in this case again show a temperature decrease in the chromosphere above the height of formation of the \ion{Ca}{2} 8542 \AA\ line (albeit at slightly lower column mass, greater height than for the hot pixels). We also note that for the IBIS and Band 6 inversions, the vertical extent of the photosphere also appears to increase, with the onset of temperatures below $\sim$ 5000 K occurring at $\mathrm{\log}\,\xi \sim -1.3$ instead of $\mathrm{\log}\,\xi \sim -0.7$ in the spectral line only inversions.
As for the hot pixels, it is not clear whether the inclusion of the Band 6 brightness temperature together with the inversion of the \ion{Ca}{2} 8542 \AA\ line leads to more realistic inferred temperature profiles.

These issues likely stem from the limitations of a hydrostatic, time-independent model atmosphere, in which the ionization fraction is calculated from the temperature and gas pressure.
This model is unable to account for phenomena such as rapid adiabatic cooling, in which the cooling timescale is significantly shorter than the recombination timescale, which can lead to pockets of cold gas with an overabundance of free electrons.
The opposite problem, in which the hotter gas of the mid- to upper-chromosphere is assumed to contribute too many free electrons, may also contribute to the poor fits. The proper distribution of free electrons along the atmospheric column is critical for determining the height sensitivity of the 1.2 mm radiation. This may lead to extended response functions, in which case the observed radiation temperatures become a weighted average across a range of heights that may extend from the temperature minimum to the upper-chromosphere. This temperature averaging, and its sensitivity to the time-dependent electron density, may make it critically challenging for inversion processes to properly interpret the observed $T_b$.
This may not be possible with current non-LTE inversion codes, and may require either an independent measurement of the electron density at various heights, or more likely, an inversion method that does not rely on time-independent hydrostatic equilibrium.

It should be noted that these shortcomings in our inversions are not necessarily due solely to the ALMA Band 6 observations.
Recently, \citet{2020A&A...634A..56D} performed spectroscopic inversions on IRIS observations of a plage in the \ion{Mg}{2} h and k lines, together with simultaneous observations in ALMA Band 6.
They found that these observations synergize well together, with the Band 6 continuum effectively constraining temperatures in the low- to mid-chromosphere, slotting in neatly between the wings and core of the \ion{Mg}{2} lines.
Also noteworthy is that these joint IRIS and ALMA inversions obtained some extremely cold chromospheres in certain regions; however, in that case, the cold chromospheres were located in quiet regions corresponding to extremely low Band 6 temperatures, whereas some of our joint inversions find cold chromospheric regions even in locations with moderate Band 6 temperatures in the middle of the plage (Figure \ref{fig:nlte_alma6_comp}). It may be that inversions combining spectral lines that form at overlapping heights to the millimeter radiation, and whose information might conflict under the simplified assumptions of the inversion code, lead to solutions that are not representative of the true, dynamic solar chromosphere.
Further work is needed to understand the effective source regions of the Band 6 radiation and how to properly integrate it with other spectral lines.

\subsection{Simultaneous inversions with both ALMA bands}
\label{sec:both_ALMA_inversions}

\begin{figure}
    \centering
    \includegraphics[width=0.8\textwidth]{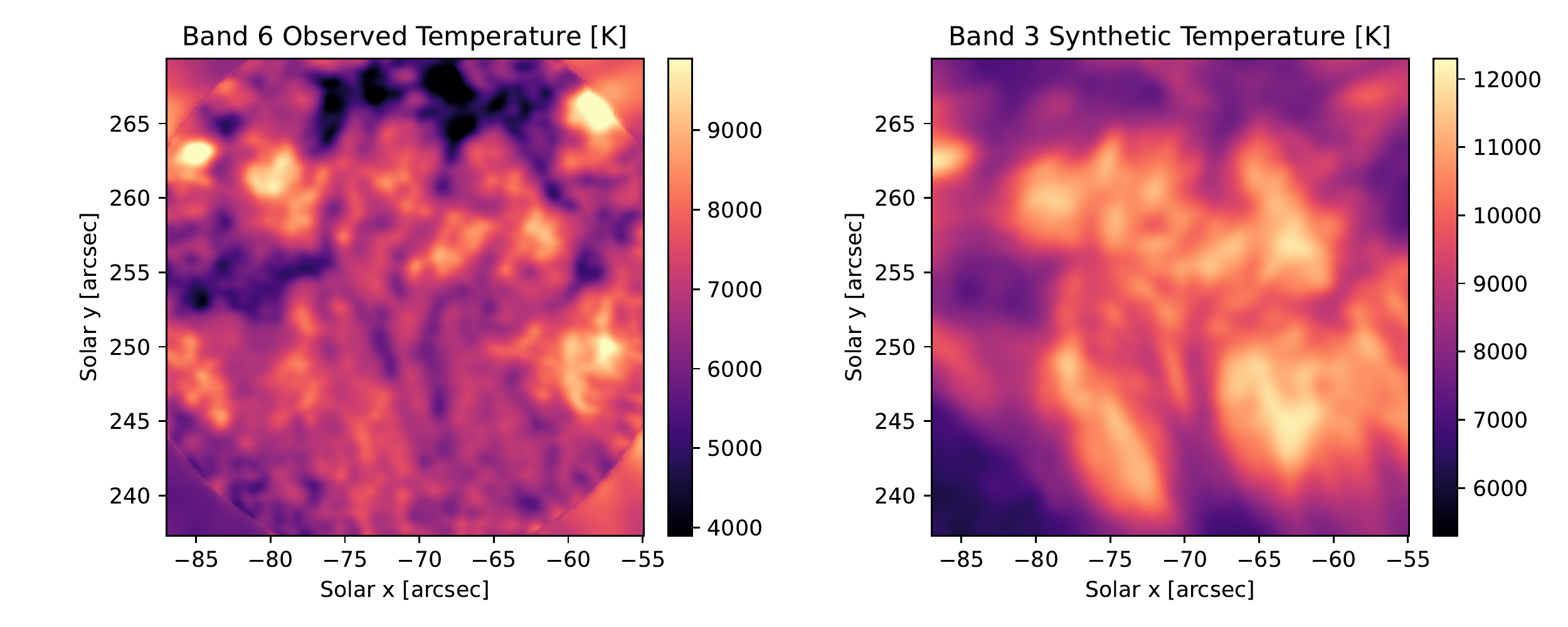}
    \caption{Millimeter continuum images used for combined inversions of series 1554 data. 
    \textbf{Left:} observed Band 6 continuum; \textbf{right:} synthetic Band 3 continuum derived from cotemporaneous H-alpha line width measurement.}
    \label{fig:band6_vs_band3}
\end{figure}

\begin{figure}
    \centering
    \includegraphics[width=0.8\textwidth]{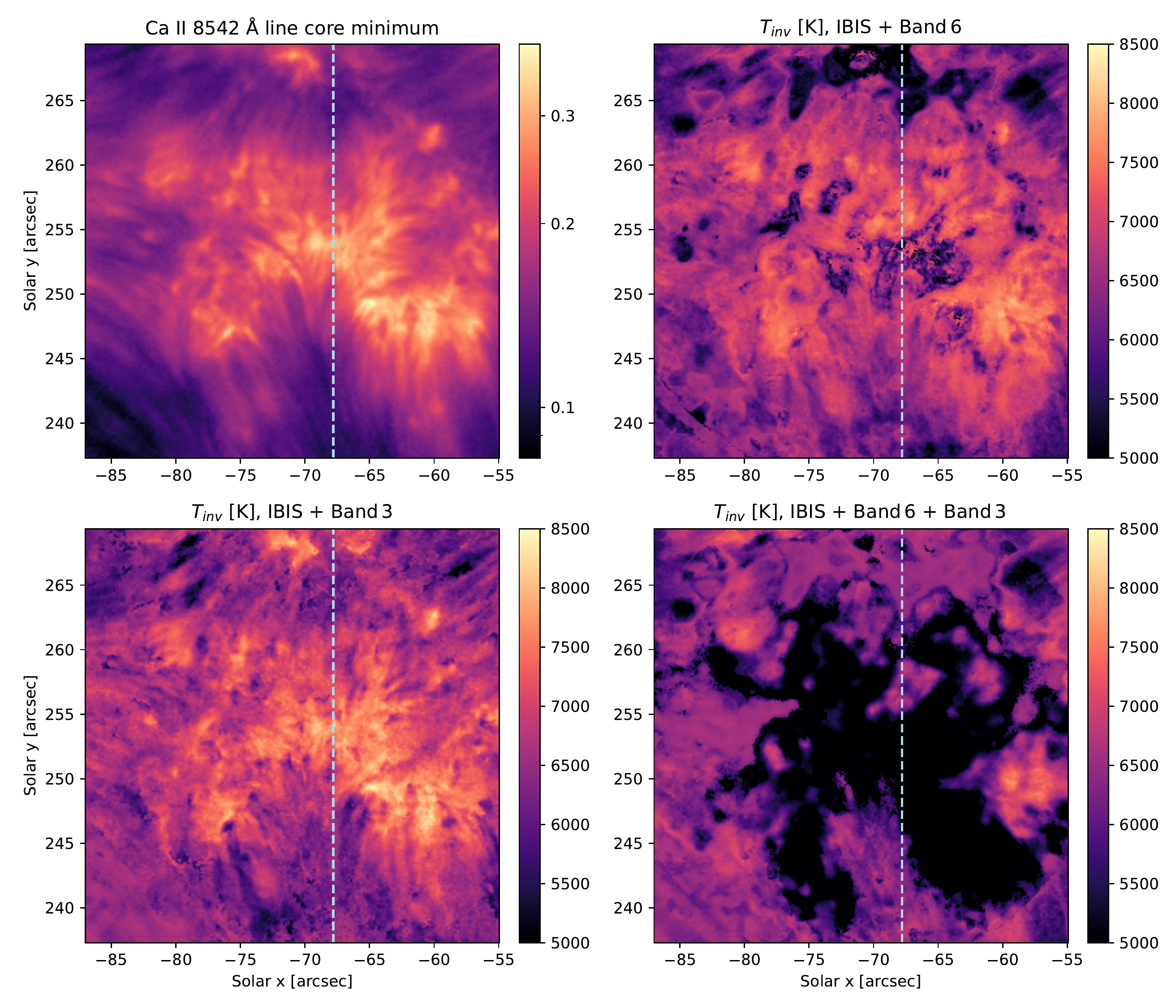}
    \caption{Temperatures in the chromosphere ($\log \xi = -3.8$), as inferred from inversions of the series 1554 data that include different combinations of millimeter continuum as input. Full stratifications along the dashed line are shown in Figure \ref{fig:ca_band6_temps_col200}. \textbf{Top left:} logarithm of the minimum intensity in the core of \ion{Ca}{2} 8542 \AA\ line.
    }
    \label{fig:ca_band6_temps}
\end{figure}

\begin{figure}
    \centering
    \includegraphics[width=0.8\textwidth]{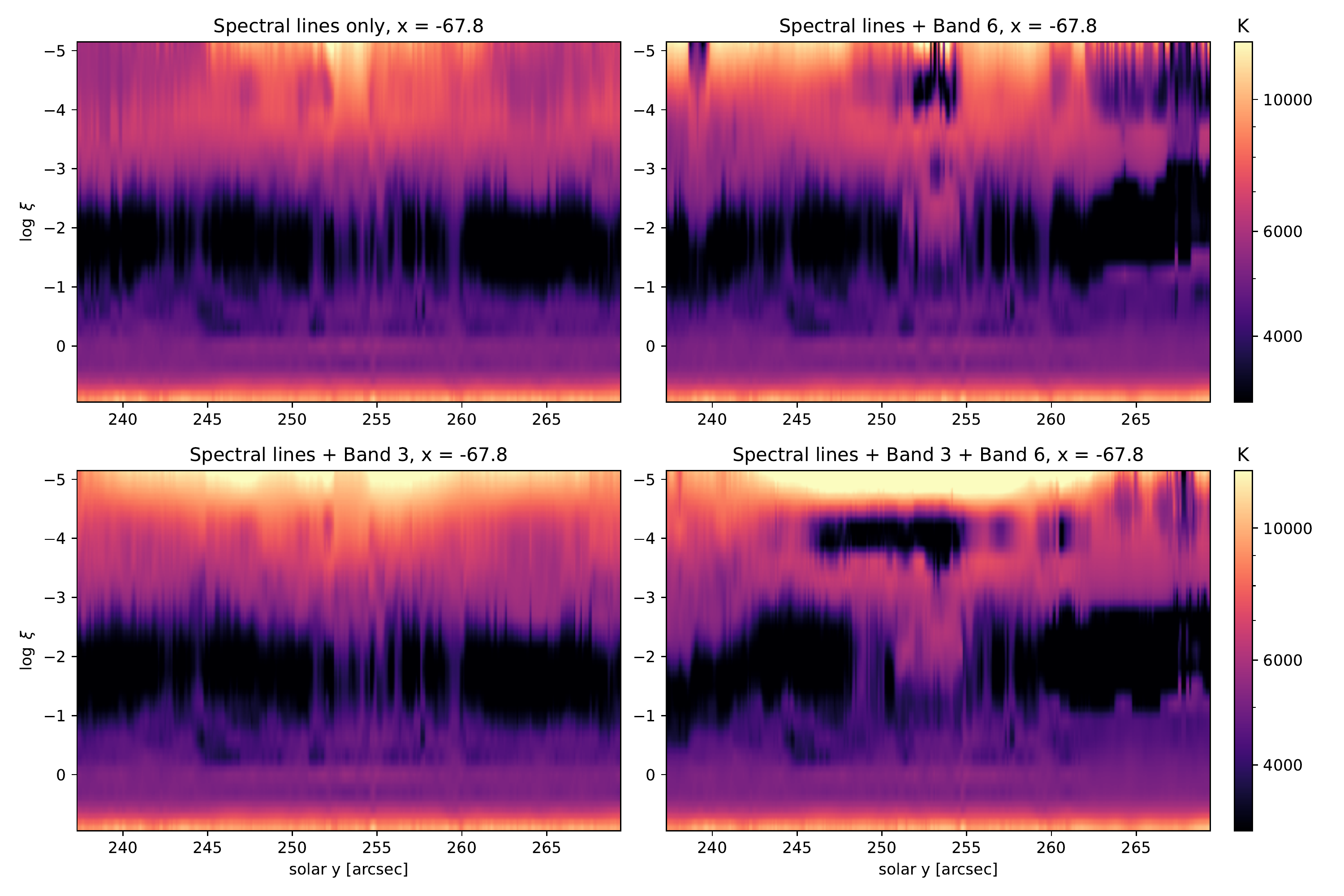}
    \caption{Temperature stratifications along the line $X = -68$ in Figure \ref{fig:ca_band6_temps}, showing the effects of including different millimeter continuum maps in the inversions. \textbf{Top left:} no millimeter inputs; \textbf{top right:} Band 6 only; \textbf{lower left:} Band 3 proxy only; \textbf{lower right:} Band 6 and Band 3 proxy. The two inversions including Band 6 result in extended areas of low temperatures high in the chromosphere.
    }
    \label{fig:ca_band6_temps_col200}
\end{figure}

This dataset provides us the opportunity to investigate the value of employing simultaneous observations of both Band 3 and Band 6 wavelengths in the inversions with the goal of better constraining the chromospheric temperature profile. Obtaining such a dataset is not possible with the currently available observing modes with ALMA, but has been discussed as a possible development of the facility capabilities, in essence dedicating subsets of the array antennas to different wavelength bands. In order to simulate such as dataset, we make use of the linear correlation between the ALMA Band 3 brightness temperatures and the H$\alpha$ line widths \citep{2019ApJ...881...99M}. We take the H-alpha width map obtained simultaneously with the actual Band 6 observation, smooth it with a 2\arcsec\ FWHM Gaussian smearing to match the ALMA resolution and apply their linear fit parameters to generate a corresponding Band 3 temperature map, shown in Figure \ref{fig:band6_vs_band3}. The simulated Band 3 image looks roughly similar to that observed 90 minutes later during series 1725 (compare to Figure \ref{fig:inv_vs_alma_band3}), with changes consistent with the typical evolution on these time scales. 

The two temperature maps show similarities in some of the spatial structuring, most notably in the regions where the brightness temperatures are either high ($T_b^{1.2mm} > 8000 K$) and low ($T_b^{1.2mm} < 5000 K$). In the regions with the intermediate temperatures, the correspondence between the two maps is much less pronounced. This might indicate that in hotter regions, the height of formation of the 1.2\,mm continuum increases to more closely match that of the 3\,mm wavelengths, and that conversely in colder regions the height of formation of the 1.2\,mm interval might move lower in the atmosphere. The differences in the appearance in the intermediate regime might indicate that the heights of formation of the continuum at these two wavelengths might not always track as closely as suggested by \citep{2020ApJ...891L...8M}.

In order to provide a direct comparison with the inversions incorporating the Band 6 data, using the same spectral line scans and solar structures, we used the proxy Band 3 temperature maps as an input into the inversion of the series 1554 data. The recovered temperatures at log $\xi$ = -3.8, corresponding most closely to the height of formation of the core of the \ion{Ca}{2} 8542 \AA\ line, is shown in the lower-left panel of Figure \ref{fig:ca_band6_temps}. Similarly to the results in Section \ref{sec:ALMA3_and_spectra_inversions}, including the Band 3 radiation alone doesn't significantly change the temperatures in the mid-chromosphere, which remain determined primarily by the information from the \ion{Ca}{2} 8542 \AA\ line. 

The comparison for the temperatures recovered at this height with the spectral inversions incorporating only Band 6 is shown in the upper-right of Figure \ref{fig:ca_band6_temps}. In this case the method infers regions of very cold gas both in places with low Band 6 brightness temperature (e.g. at the top of the field), but also in the hotter central regions above the magnetic network. Both of these inversion-derived temperatures can be compared to the intensity in the core of the \ion{Ca}{2} 8542 \AA\ line (upper-left) or to the temperatures derived at this same height when carrying out the inversions only the spectral lines (left panel of Figure \ref{fig:inv_vs_alma_band3}).

Finally, we performed the same inversions using both the observed Band 6 and proxy Band 3 images as additional inputs into the inversions, with the derived temperatures at this same height shown in the bottom right of Figure \ref{fig:ca_band6_temps}. Compared to the inversions with just one of the ALMA bands, these joint inversions achieved overall worse fits and display similar issues over a wider physical region.
In particular, the regions of extremely cold gas noted in section \ref{sec:ALMA6_and_spectra_inversions} now extend across much of the plage, while for the regions with particularly low Band 6 temperatures, many of the inversions were unable to reproduce the observed Band 6 temperatures.

To understand the nature of the solutions found with the different combinations, we examine the temperature profiles for a slice through the data cube (whose locations is indicated by the white dashed line in Figure \ref{fig:ca_band6_temps}. The temperatures inferred in the case of using only the spectral lines are shown in the upper left of Figure \ref{fig:ca_band6_temps_col200}, while the other three panels show combinations that include one or the other ALMA bands, or both (lower right). The spectral-line-only inversions show a relatively well behaved rise in temperatures from T $\sim$ 4000 K in the upper photosphere to T $\ge$ 7000 K in the chromosphere. However, only with the inclusion of the Band 3 temperatures do we see a strong rise at the top of the inversion boundary, with T $\ge$ 11000 K at log $\xi\,\le$\,-5. Otherwise, the inclusion of the Band 3 information doesn't produce significant effects in the photosphere or low chromosphere.

However, when using only the Band 6 as an additional input into the spectral inversions, certain recovered atmospheric profiles become problematic, with very low temperatures T $<$ 4000 K found in the mid chromosphere. It also changes the inferred conditions lower down, with high temperatures extending down into the nominal temperature minimum (251\,$<$\,y\,$<$\,255) or moving the temperature minimum region higher(265\,$<$\,y\,$<$\,269). This does produce some atmospheres that have a strong temperature rise at log$\xi$ = -5, but many some display surprisingly low temperatures at the top of the boundary.

When including both Band 3 and Band 6 as an input, the temperature rise at the top of the boundary is more uniform, but the regions of low temperatures in the middle chromosphere (log$\xi$ = -4) are even more extended (as was seen in the spatial maps in Figure \ref{fig:ca_band6_temps}. As a reminder, this portion of the slice lies above a region of relatively enhanced magnetic field and bright \ion{Ca}{2} 8542 \AA\ line core intensity.

Overall, these different combinations of the inversions reinforce what we had shown earlier. The addition of (proxy) Band 3 to the inversion of the spectral lines primarily enforces solutions showing a strong temperature rise above the height of formation of the core of the \ion{Ca}{2} 8542 \AA\ line, as is physically reasonable. The inversions including Band 6 can be very unreliable in certain pixels, in a way that is not easy to predict. These problems are only exacerbated when including both ALMA wavelengths at the same time. This appears to create solutions, with pockets of very cold plasma in the middle chromosphere, that seem to be unrealistic and don't correspond to the signal from \ion{Ca}{2} 8542 \AA\ expected to form in this same region.

\section{Discussion and Conclusions} 
\label{Ch:Conclusion}

We have presented observations of a solar plage region observed simultaneously in two visible/near-IR spectral lines (\ion{Ca}{2} 8542 \AA\ and \ion{Na}{1} 5896 \AA) and in the millimeter continuum.
We performed spectroscopic inversions on these data to recover the temperature profiles through the photosphere and chromosphere at each spatial pixel.
For a typical inversion setup, when using a fixed-number of regularly space nodes, we found that using column mass for the height scale in the inversion, instead of optical depth, was more reliable in inferring the chromospheric physical parameters.
This is because the column mass scale naturally spreads the chromospheric temperature rise over a greater range of nodes in this height scale, which can more easily and stably be reproduced in the inversion process.
Even though the column mass scale tends compress the photosphere into a smaller range of column-mass values ($-2 < \log\xi < 1$ versus $-4 < \log\tau < 0$), we found it was still reliable in fitting the wings of the lines and recovering the conditions in the lower atmosphere.
In fact, our inversion-derived temperature profiles in the photosphere using the optical depth scale showed small-scale fluctuations with height that were not present in the column-mass inversions.
Similarly, a dip in the temperature profiles in the upper chromosphere (at the limit of the \ion{Ca}{2} 8542 \AA\ sensitivity was present in the optical-depth inversions, while in most cases the column-mass inversions showed a nearly monotonic temperature rise at those heights.

We then compared the temperatures derived from the joint inversion of the \ion{Ca}{2} 8542 and \ion{Na}{1} 5896 \AA\ lines (using the column-mass height scale) with the simultaneous brightness temperatures measured by ALMA at 1.25 and 3.0 mm. 
These two, independent temperature measurements show some correlation, as expected.
The Band 3 images exhibit similar large-scale structures as the temperature maps from the approximate height of formation of the \ion{Ca}{2} 8542 line core. The correspondence breaks down at smaller scales around $<\sim 5 \arcsec$ (more than twice the $\sim 2 \arcsec$ beam size). In addition, the observed $T_b^{3mm}$ is higher than the temperatures inferred for the \ion{Ca}{2} 8542 \AA\ line core ($\sim 9500 \text{K}$ vs. $\sim 7000 \text{K}$); One explanation is that the 3-mm continuum forms, on average, somewhat  higher in the chromosphere than the \ion{Ca}{2} 8542 \AA\ line and is thus sensitive to the rising temperatures at the base of the transition region \citep[similar to the case for H$\alpha$ discussed by][]{2019ApJ...881...99M}.

By contrast, the Band 6 temperature maps exhibit significant structural differences from the temperatures inferred from the inversion of the \ion{Ca}{2} 8542 line, even if these two diagnostics could be expected to form in more similar heights in of chromosphere. 
In fact, there was no layer in the temperature cube recovered by the inversions that showed a good spatial correlation with the 1.2 mm continuum image. 
Curiously, the mean temperature over the field view  ($\sim 7000 \text{K}$) was very similar for the Band 6 and for the inversions at the layer best corresponding to the \ion{Ca}{2} 8542 line core. 
This correspondence of the mean temperatures but bad spatial correlation may indicate that the temperatures observed at 1.2 mm might be an average over a large spread of heights in the atmosphere that show different spatial structuring, but scattered around a common average temperature. 
Previous works have already suggested that Band 6 could have a disjoint contribution function with components at various heights \citep[][]{2015A&A...575A..15L, 2021ApJ...906...83C}.
The observed Band 6 continuum emission at any given resolution element thus might include contributions from multiple different plasma elements along the line of sight, leading to a local breakdown of the correlation with the \ion{Ca}{2} 8542 \AA\ line, which is formed over a narrower range of heights. 
This may be especially problematic in the current situation, where the spatial resolution in the millimeter observations is lower than in the optical and IR lines.
Alternatively, this might indicate an offset of unknown source in the absolute temperatures derived by the inversions or the ALMA calibration \citep{2019A&A...622A.150J}. 
We also compared the appearance of the Band 6 and Band 3 observations of the same region taken only 90 minutes apart. The differences in the observed structures seem to be greater than what would be expected from solar evolution or due to the different spatial resolution, which isn't fully consistent with the result from non-equilibrium-ionization simulations that these two continuum emissions are generally formed at the same heights in the atmosphere \citep{2020ApJ...891L...8M}.

We further investigated the relationship between ALMA observations and spectral line diagnostics by performing joint inversions, in which the observed millimeter $T_b$ were included as an additional constraint in the fitting process.
Inclusion of ALMA Band 3 data provided a seemingly reliable constraint on the temperatures inferred in the upper chromosphere, ensuring that there was a strong temperature rise at the top of the atmospheric range. 
Without the inclusion of the Band 3 temperatures, the recovered atmospheric profiles tended to remain flat or even go down slightly about the primary height of formation of the \ion{Ca}{2} 8542 line core. 
Otherwise, the inclusion of Band 3 in combination with the two spectral lines had little effect on the inferred temperature stratifications lower down.

Inclusion of Band 6 data, on the other hand, drastically alters the inferred atmospheres, introducing locations with temperatures at chromospheric heights as low as 4000 K above the plage. 
The inversion also found regions of lower temperatures also lower down in the atmosphere, but generally away from the strong magnetic concentrations.
We spatially partitioned the $T_b^{1.2mm}$ temperature map into regions of high ($T_b > 7900 K$) or low ($T_b < 5900 K$) temperatures. 
We found good spatial correlations for the regions of hotter temperatures with the corresponding inversion-derived temperatures above the height of formation of the \ion{Ca}{2} 8542 line ($\log\xi = -4.4)$) and for the lower temperatures regions with the temperatures in those spatial locations but much deeper in the atmosphere ($\log\xi = -2.9)$, just above the temperature minimum). 
In the regions of intermediate temperatures, there was again no height that showed a strong correlation between these two temperature measures. 
This was shown to be consistent with the changes in the distribution of heights for the response functions we calculated for Band 6 (and to some extent Band 3), ranging from $\log\xi = -1.5$ for the coldest pixels to $\log\xi < -4$ for the hottest.
Even attempting to perform an inversion with the two spectral lines and both Band 6 and Band 3 brightness temperatures (the estimate for the latter being derived from simultaneous H$\alpha$ data using the relationship found by \cite{2019ApJ...881...99M} didn't improve the reliability of the inversions.
It appears that the inversion process has some difficulty in determining the proper height of formation of the 1.2 mm continuum in any given pixel.

This is likely related to another issue we examined in this work, the need to include the option in the STiC code to treat non-LTE hydrogen ionization in the process of solving for the underlying atmospheric conditions (and in particular the electron pressure). 
The effect of including this mechanism is to reduce the hydrogen ionization, and hence the electron density, for a given temperature or height in the chromosphere. 
This reduced density also manifests itself in a reduced collisional coupling between the \ion{Ca}{2} 8542 source function and the local temperature. 
To reproduce the observed line core intensities for this line, the inversion process must infer a higher local temperature. 
Therefore, when we included this non-LTE ionization in the inversions, we recovered higher temperatures in the chromosphere heights corresponding to the \ion{Ca}{2} 8542 line, more closely matching the temperatures measured in the 3 mm continuum with ALMA. 
However, even including this effect, it seemed the inversion fitting process struggled to reconcile the temperatures measured in the 1.2 mm continuum in particular, with the temperatures needed to recreate the observed spectral lines forming at similar heights.

The underlying issue may be that both of these diagnostics are somewhat sensitive to the electron density profiles through the atmosphere. 
For the \ion{Ca}{2} 8542 line this is through the collisional coupling described above, while for the millimeter continuum the density strongly controls the contribution height(s) of the continuum emission.
With a strong gradient in temperature with height, uncertainties in the electron density can lead to significant confusion in the inversion.
Uncertainties in understanding the electron density profile in the solar atmosphere are a known problem, as indicated by the discrepancies in this parameter for the FAL-like models based on spectral-line intensities and those models calculated to explicitly fit the limb intensities in the radio and millimeter wavelengths \citep{Ewell_CICM_1993, Selhorst_SSC_2005, Selhorst_2019}. The latter show elevated electron densities extending several thousand kilometers higher in the atmosphere.   
This difficulty in understanding this component of the atmospheric structure is exacerbated by the fact that the calculations underlying the inversion and line fitting process do not strongly constrain the determination of the actual electron density profile, since variations can be accommodated by changes in the temperature profile in order to achieve a reasonable fit. 
Further, the electron density depends on complicated physical processes that control the hydrogen ionization \citep[][]{2017IAUS..327....1R}, which are hard to fully incorporate in a computationally tractable manner. Further developments in this area could include a better treatment of the electron density in order to better reconcile disparate sources of temperature measurements in the chromosphere.

\begin{acknowledgements}
    {The authors would like to express their gratitude for the support staff at the DST, especially Doug Gilliam, the observing staff at ALMA, especially Juan Cortes, and the ALMA support scientist, Tim Bastian.} 
    
    {The National Solar Observatory (NSO) is operated by the Association of Universities for Research in Astronomy, Inc. (AURA), under a cooperative agreement with the National Science Foundation.}
    {IBIS has been designed and constructed by the INAF/Osservatorio Astrofisico di Arcetri with contributions from the Universit\`a di Firenze, the Universit\`a di Roma ``Tor Vergata'', and upgraded with further contributions from NSO and Queens University Belfast.}
    
    {This paper makes use of the following ALMA data: ADS/NRAO.ALMA\#2016.1.01129.S. ALMA is a partnership of ESO (representing its member states), NSF (USA) and NINS (Japan), together with NRC (Canada), MOST and ASIAA (Taiwan), and KASI (Republic of Korea), in cooperation with the Republic of Chile. The Joint ALMA Observatory is operated by ESO, AUI/NRAO and NAOJ. }
    {The National Radio Astronomy Observatory is a facility of the National Science Foundation operated under cooperative agreement by Associated Universities, Inc.}

    {RH was supported by the DKIST Ambassador Program  as well as with funding provided though NASA Grant 80NSSC20K0179. 
    Funding for the DKIST Ambassadors program is provided by the National Solar Observatory, a facility of the National Science Foundation, operated under Cooperative Support Agreement number AST-1400405.
    Support for the work of RH was also provided by the NSF through award SOSPA6-022 from the NRAO.}
    {KPR acknowledges the support of NASA under the grant 80NSSC20K1282.
    MEM was supported by the George Ellery Hale Graduate Student Fellowship from the University of Colorado, Boulder.}
    {The authors would like to thank Gianna Cauzzi for valuable comments on the manuscript.}
    
    {This work utilized resources from the University of Colorado Boulder Research Computing Group, which is supported by the National Science Foundation (awards ACI-1532235 and ACI-1532236), the University of Colorado Boulder, and Colorado State University. }
    {This research has made use of NASA's Astrophysics Data System as well as Rob Rutten's SDO alignment IDL package.
    This research utilized the Python libraries matplotlib \citep{matplotlib} and the NumPy computational environment \citep{numpy}.}
\end{acknowledgements}

\facility{ALMA, Dunn (IBIS), SDO (AIA, HMI), IRIS}
\pagebreak

\bibliography{alma_inversion_paper.bib}{}

\end{document}